\documentclass[12pt]{article} 
\pdfoutput=1
\usepackage{jheppub}
\newcommand{\MET}{E\llap{/\kern1.5pt}_T}
\usepackage{amsmath}
\usepackage{amsfonts}
\usepackage{bbold}
\usepackage{epsfig}
\title{A weakly constrained $W^{\prime}$ at the early LHC} 
\author[a,b]{Christophe Grojean}
\author[a,c]{Ennio Salvioni}
\author[a,d,e]{Riccardo Torre}
\affiliation[a]{CERN, Physics Department, Theory Division, CH-1211 Geneva 23, Switzerland}
\affiliation[b]{Institut de Physique Th\'eorique, CEA Saclay, F-91191 Gif-sur-Yvette C\'edex, France}
\affiliation[c]{Dipartimento di Fisica and INFN, Universit\`a di Padova, Via Marzolo 8, I-35131 Padova, Italy}
\affiliation[d]{Dipartimento di Fisica and INFN, Universit\`a di Pisa, Largo Fibonacci 3, I-56127 Pisa, Italy}
\affiliation[e]{Institut f\"ur Theoretische Physik, Universit\"at Z\"urich, Winterthurerstrasse 190, CH-8057 Z\"urich, Switzerland}
\emailAdd{Christophe.Grojean@cern.ch}
\emailAdd{Ennio.Salvioni@cern.ch}
\emailAdd{Riccardo.Torre@cern.ch}
\proceeding{\vspace{-1.65cm}\flushright{\small CERN-PH-TH/2011-041\\
DFPD-2011/TH/3\\
IFUP-TH/2011-05\\
LPN11-10\\
ZU-TH 04/11\\}}
\abstract{We study, within an effective approach, the phenomenology of a charged $W'$ vector which transforms as an isosinglet under the Standard Model gauge group. We discuss bounds from present data, finding that these are quite weak for suitable choices of the right-handed quark mixing matrix. Then we study the resonant production at the early LHC of such a weakly constrained $W'$. We start discussing the reach in the dijet final state, which is one of the channels where the first $W'$ signal would most likely appear, and then we analyse prospects for the more challenging discovery of $W'$ decays into $W\gamma$ and $WZ$. We show in particular that the former can be used to gain insight on the possibly composite nature of the resonance.  
}
\begin{document} 
\maketitle
\section{Introduction and theoretical motivations}

Heavy spin-1 resonances are a generic prediction of many Beyond-the-Standard Model (BSM) theories. The most frequently discussed case is that of new gauge bosons associated with extensions of the SM gauge group. While neutral states, known in the literature as $Z'$ \cite{Langacker:2008yv,Salvioni:2009p068,Salvioni:2010p010,Accomando:2010fz}, can be introduced by simply adding an extra $U(1)$ factor to the SM gauge group $G_{\mathrm{SM}}=SU(3)_{C}\times SU(2)_{L}\times U(1)_{Y}$, electrically charged states need a non-Abelian extension of the SM gauge symmetry. Some well-known examples of such extensions are those appearing in grand unified theories, including Left-Right (LR) models, in Little Higgs models, and in models where the Higgs is a pseudo-Goldstone boson arising from the spontaneous breaking of an extended global symmetry. A $W'$ can also appear as a Kaluza-Klein excitation of the $W$ in theories with extra dimensions.

On the other hand, there is also the interesting possibility that such heavy spin-1 particles are composite states, bound by a new strong interaction responsible for ElectroWeak Symmetry Breaking (EWSB). The most convenient approach in the discussion of the LHC reach on such composite resonances is to write the most general effective Lagrangian describing interactions of the new state with the SM fields and invariant under $G_{\mathrm{SM}}$ (see, e.g., Refs.~\cite{Bauer:2010p280,Barbieri:2011p2759,Han:2010p2696}). Once the representation in which the extra state transforms is specified, the Lagrangian is fully determined by a set of free parameters, namely the mass of the heavy state and its couplings to the SM particles. A specific gauge model, in which the vector is the gauge boson associated with the gauging of some extra symmetry, can then be recovered by taking some special values of these free parameters. 

We apply an effective approach to study the early LHC phenomenology of a $W'$ transforming in the representation 
\begin{equation} \label{111}
(\mathbf{1},\mathbf{1})_{1}
\end{equation}
of $G_{\mathrm{SM}}$, where the notation $(SU(3)_{c},SU(2)_{L})_{Y}$ has been adopted. A similar approach has been employed by the authors of Ref.~\cite{Aguila:2010p1781}, where however the focus was on computing constraints from electroweak data. In Ref.~\cite{Aguila:2010p1781}, bounds from ElectroWeak Precision Tests (EWPT) were discussed for all the irreducible representations of the SM gauge group which can have linear and renormalizable couplings to SM fields. There it was shown that the only such representations containing a color-singlet $W'$ (for a study of colored resonances at the early LHC, see Ref.~\cite{Han:2010p2696}) coupled to the SM fermions, in addition to that in Eq.~\eqref{111}, are $(\mathbf{1},\mathbf{3})_{0}$ and $(\mathbf{1},\mathbf{2})_{-3/2}$. The $(\mathbf{1},\mathbf{2})_{-3/2}$ multiplet does not have any renormalizable coupling to quarks or gluons, and as a consequence its production at the LHC would be very suppressed: therefore, we do not discuss it any further in the present work. Our choice to discuss the representation $(\mathbf{1},\mathbf{1})_{1}$ is motivated by the fact that in this case we can add to the SM only a charged resonance, without any associated neutral state. This is in contrast with the other representation commonly obtained in specific models, namely the $SU(2)_{L}$ triplet $(\mathbf{1},\mathbf{3})_{0}$. In the latter case, the $W'$ and $Z'$ masses are degenerate, apart from electroweak scale corrections, and as a result the strong bounds from neutral currents (including LEP2 data on four-fermion operators) apply also to the $W'$, pushing its mass well into the TeV range (and thus out of the LHC reach in its first run) unless its couplings to leptons are very small. On the other hand, a $W'$ transforming as $(\mathbf{1},\mathbf{1})_{1}$, because its only couplings to leptons arise through $W$-$W'$ mixing and are therefore strongly suppressed\footnote{Since we only consider the SM field content, we do not include right-handed neutrinos; or, equivalently for our purposes, we assume them to be heavier than the $W'$, so that the decay $W'\rightarrow \ell_{R}\nu^{\ell}_{R}$ is forbidden.}, is only constrained by hadronic processes (except for the oblique $T$ parameter). As we will discuss later, if particular forms for the right-handed quark mixing matrix are chosen as to evade constraints from $\Delta F=2$ transitions, the coupling of the $W'$ to quarks is only constrained by Tevatron direct searches, and therefore it can be sizable, without violating any existing constraint, even for a $W'$ mass below one TeV, making a discovery of the resonance at the early LHC possible. Furthermore, as discussed in Refs.~\cite{Hsieh:2010p2579,Schmaltz:2010p2610}, in Left-Right (LR) models, which give a $(\mathbf{1},\mathbf{1})_{1}$ charged state after LR symmetry breaking, the splitting between the masses of the $W'$ and $Z'$ (with the latter being a singlet under $G_{\mathrm{SM}}$) can be large, without violating EWPT constraints, if one takes $g_{X}\gg g_{R}$, where $g_{X}$ and $g_{R}$ are the couplings of the Abelian factor and of $SU(2)_{R}$, respectively. Also assigning the Higgs responsible for $SU(2)_{R}\times U(1)_{X}\rightarrow U(1)_{Y}$ breaking to a higher dimensional representation (for example, introducing a $SU(2)_{R}$ triplet Higgs) can help in increasing the mass splitting between the $W'$ and $Z'$. If such splitting is large enough, constraints from the $Z'$ can be made negligible, and one can study the phenomenology of the $W'$ using an effective theory for a $(\mathbf{1},\mathbf{1})_{1}$ state. Another example of a construction where the $W'$ we consider arises is the Littlest Higgs with custodial symmetry \cite{Chang:2003p057} (incidentally, we remark that several Little Higgs models contain in the spectrum a spin-1 $SU(2)_{L}$ triplet). While these provide specific examples of $W'$ that are described by the effective theory we consider, the interest of our approach goes much further, as it encompasses any composite state, whose properties could depart significantly from those of the gauge boson of a minimal non-abelian extension of $G_{\mathrm{SM}}$. We also note that a $W'$ with flavor-violating couplings to quarks has been invoked as an explanation of the anomaly in the top pair forward-backward asymmetry observed by CDF: we briefly comment on how such a $W'$ is described by our framework in Section~\ref{indirect bounds}. Composite vectors are usually considered in Higgsless models or in models where the Higgs is a composite state, where they have been shown to play an important role in keeping perturbative unitarity in the longitudinal $WW$ scattering up to the cut-off \cite{Csaki:2003dt,Barbieri:2008p1580}. The LHC phenomenology of these composite states is discussed, e.g., in Refs.~\cite{He:2008p3365, Barbieri:2010p144, Barbieri:2010p1577, CarcamoHernandez:2010p1578, Cata:2009p713,Birkedal:2004au,Martin:2009gi}. 

We discuss the prospects of the early LHC to discover the $W'$ in the dijet channel, which, together with the $tb$ final state \cite{Gopalakrishna:2010p2723}, is the main avenue to look for the `leptophobic' $W'$ we are considering. A particularly striking difference between gauge models and the effective theory we consider is the presence in the latter case of a sizable $W'W\gamma$ interaction, which is very suppressed if the $W'$ is a fundamental gauge boson. As a consequence, observation of the $W'\rightarrow W\gamma$ decay at the LHC would be a hint of the compositeness of the resonance. In this light, we discuss the LHC prospects for discovery of the $W'\rightarrow W\gamma$ decay. We also present the prospects for observing the $W'\rightarrow WZ$ decay at the early LHC, and compare the reach in this channel to that in the $W\gamma$ final state. For previous relevant work on the phenomenology of a $W'$ at the LHC, see Refs.~\cite{Schmaltz:2010p2610,Frank:2010p2250,Rizzo:2007p3440,Nemevvsek:2011hz}. In Ref.~\cite{Schmaltz:2010p2610}, the early LHC reach on two simple $W'$ models was discussed. Our work differs from the discussion of a right-handed $W'$ in Ref.~\cite{Schmaltz:2010p2610} in two ways: firstly, as already detailed above we adopt an effective approach, without relying on any specific model; secondly, we make the `pessimistic' assumption that the decay of $W'$ into right-handed neutrinos, which was studied in Ref.~\cite{Schmaltz:2010p2610} (see also Ref.~\cite{Nemevvsek:2011hz}), be kinematically closed, and discuss the reach in the dijet and diboson final states. 

Our paper is organized as follows: after introducing the effective Lagrangian in Section~\ref{model-indep approach}, we discuss bounds on the parameter space of the model coming from electroweak and low-energy data in Section~\ref{indirect bounds}, and from Tevatron searches in Section~\ref{Tevatron}. Section~\ref{secLHC} is devoted to the study of the early LHC reach on the $W'$ we are discussing: in Section~\ref{LHC} we present results for the dijet final state, in Section~\ref{LHCWgamma} we study the $W'\rightarrow W\gamma$ channel and we discuss how it could be used to obtain information on the theoretical nature of the resonance; the complementary search for $W'\rightarrow WZ$ is discussed in Section~\ref{WZsearch}. Finally, we present our conclusions in Section~\ref{summary}. Appendix~\ref{Partialdecaywidths} contains the partial decay widths of the $W'$, whereas in App.~\ref{EffLagr130} the effective Lagrangian for a $W'$ transforming as an $SU(2)_{L}$ triplet is given for completeness. In App.~\ref{minimalgaugemodels} we set the notation for the most economic gauge extensions of the SM containing either an iso-singlet or an iso-triplet $W'$.   


\section{`Model independent' approach} \label{model-indep approach}
We consider, in addition to the SM field content, a complex spin-1 state transforming as a singlet under color and weak isospin, and with hypercharge equal to unity, according to Eq.~\eqref{111}. The extra vector is therefore electrically charged, with unit charge (we adopt a normalization for the hypercharge such that the electric charge is $Q=T_{3L}+Y$, where $T_{3L}$ is the third component of the weak isospin). We do not make any assumption on the theoretical origin of the extra state, and in particular we do not assume it to be a gauge boson associated with an extended gauge symmetry. Taking a \mbox{model-independent} approach, we write down all the renormalizable interactions between the new vector and the SM fields which are allowed by the $SU(3)_{c}\times SU(2)_{L}\times U(1)_{Y}$ gauge symmetry. Higher-Dimensional Operators (HDO) would be suppressed with respect to renormalizable ones by the cut-off of the theory; we neglect them in our analysis. We expect HDO to give corrections roughly of order $M^{2}_{W'}/\Lambda^{2}$ to our results: in Section~\ref{LHCWgamma} we show that the cut-off always satisfies $\Lambda\gtrsim 5M_{W'}$, so we can conservatively estimate our results to hold up to 10 percent corrections due to HDO. Within this framework, we write down the Lagrangian     
\begin{equation}  \label{lagrangian}
\mathcal{L}=\mathcal{L}_{SM}+\mathcal{L}_{V}+\mathcal{L}_{V-SM}\,,
\end{equation}
where $\mathcal{L}_{SM}$ is the SM Lagrangian, and\footnote{To be general, we should also include the operators $V_{\mu}^{+}V^{+\mu}V_{\nu}^{-}V^{-\nu}$ and $V_{\mu}^{+}V^{-\mu}V_{\nu}^{+}V^{-\nu}$; however, these operators only contribute to quartic interactions of vectors and can thus be neglected for the scope of this study. On the other hand, a cubic self-interaction of $V_{\mu}$ is forbidden by gauge invariance.}
\begin{align}
\mathcal{L}_{V}\,=\,& \nonumber D_{\mu}V_{\nu}^{-}D^{\nu}V^{+\mu}-D_{\mu}V_{\nu}^{-}D^{\mu}V^{+\nu}+\tilde{M}^{2}V^{+\mu}V_{\mu}^{-}\\ 
&+\frac{g_{4}^{2}}{2}|H|^{2}V^{+\mu}V_{\mu}^{-}-ig_{B}B^{\mu\nu}V^{+}_{\mu}V^{-}_{\nu}\,,\\ \label{V-SM}
\mathcal{L}_{V-SM}\,=\,& V^{+\mu}\left(ig_{H}H^{\dagger}(D_{\mu}\tilde{H})+ \frac{g_{q}}{\sqrt{2}}(V_{R})_{ij}\overline{u_{R}^{i}}\gamma_{\mu}d_{R}^{j}\right) +\text{h.c.}\,, 
\end{align}
where we have denoted the extra state with $V^{\pm}_{\mu}$, and have defined $\tilde{H}\equiv i\sigma_{2}H^{\ast}$. We remark that we have not introduced right-handed neutrinos, in order to avoid making any further assumptions about the underlying model. The coupling of $V_{\mu}$ to left-handed fermionic currents is forbidden by gauge invariance. The covariant derivative is referred to the SM gauge group: for a generic field $\mathcal{X}$, neglecting colour we have
\begin{equation}
D_{\mu}\mathcal{X}=\partial_{\mu}\mathcal{X}-igT^{a}\hat{W}^{a}_{\mu}\mathcal{X}-ig'YB_{\mu}\mathcal{X}\,,
\end{equation}
where $T^{a}$ are the generators of the $SU(2)_{L}$ representation where $\mathcal{X}$ lives, and we have denoted the $SU(2)_{L}$ gauge bosons with a hat, to make explicit that they are gauge (and not mass) eigenstates. In fact, upon electroweak symmetry breaking the coupling $g_{H}$ generates a mass mixing between $\hat{W}_{\mu}^{\pm}$ and $V_{\mu}^{\pm}$. This mixing is rotated away by introducing mass eigenstates 
\begin{equation}
\begin{pmatrix} W^{+}_{\mu} \\
W^{\prime\,+}_{\mu} \end{pmatrix} = \begin{pmatrix}
\cos\hat\theta\,\,\, & \sin\hat\theta \\
-\sin\hat\theta\,\,\, & \cos\hat\theta 
\end{pmatrix} \begin{pmatrix}
\hat{W}_{\mu}^{+} \\
V_{\mu}^{+}
\end{pmatrix}\,.
\end{equation} 
The expression of the mixing angle is
\begin{equation}
\tan(2\hat{\theta})=\frac{2\Delta^{2}}{m^{2}_{\hat{W}}-M^{2}}\,,
\end{equation}
where 
\begin{equation}
m_{\hat{W}}^{2}=\frac{g^{2}v^{2}}{4},\quad \Delta^{2}=\frac{g_{H}gv^{2}}{2\sqrt{2}},\quad M^{2}=\tilde{M}^{2}+\frac{g_{4}^{2}v^{2}}{4}\,.
\end{equation}
We denote with $v\approx 246$ GeV the SM Higgs \textsc{vev}. We assume that Eq.~\eqref{lagrangian} is written in the mass eigenstate basis for fermions. We have written the heavy vector mass explicitly: the details of the mass generation mechanism will not affect our phenomenological study, as long as additional degrees of freedom possibly associated with such mechanism are heavy enough. We assume that the standard redefinition of the phases of the quark fields has already been done in $\mathcal{L}_{SM}$, thus leaving only one CP-violating phase in the Cabibbo--Kobayashi--Maskawa (CKM) mixing matrix $V_{CKM}$. The \mbox{right-handed} mixing matrix $V_{R}$ does not need to be unitary in the framework we adopt here: it is in general a complex $3\times 3$ matrix. This is a relevant difference with respect to LR models, where $V_{R}$ must be unitary, as a consequence of the gauging of $SU(2)_{R}$. We normalize $g_{q}$ in such a way that $|\det (V_{R})|=1$ (a generalization of this condition can be applied if $V_{R}$ has determinant zero).

In the mass eigenstate basis both for spin-$1/2$ and spin-1 fields, the charged current interactions for quarks read:  
\begin{equation}
\mathcal{L}^{q}_{cc}\,=\,W^{+}_{\mu}\,\overline{u}^{i}\left(\gamma^{\mu}v_{ij}+\gamma^{\mu}\gamma_{5}a_{ij}\right)d^{j}+
W^{\prime\,+}_{\mu}\,\overline{u}^{i}\left(\gamma^{\mu}v^{\prime}_{ij}+\gamma^{\mu}\gamma_{5}a^{\prime}_{ij}\right)d^{j}+\text{h.c.}\,,
\end{equation}
where $u^{i}, d^{j}$ are Dirac fermions, and the couplings have the expressions
\begin{align} \nonumber
v_{ij}=& \frac{1}{2\sqrt{2}}\left(g_{q}\sin\hat\theta(V_{R})_{ij}+g\cos\hat\theta(V_{CKM})_{ij}\right)\,, \\
a_{ij}=& \frac{1}{2\sqrt{2}}\left(g_{q}\sin\hat\theta(V_{R})_{ij}-g\cos\hat\theta(V_{CKM})_{ij}\right)\,, \nonumber\\
v^{\prime}_{ij}=& \frac{1}{2\sqrt{2}}\left(g_{q}\cos\hat\theta(V_{R})_{ij}-g\sin\hat\theta(V_{CKM})_{ij}\right)\,, \nonumber\\
a^{\prime}_{ij}=& \frac{1}{2\sqrt{2}}\left(g_{q}\cos\hat\theta(V_{R})_{ij}+g\sin\hat\theta(V_{CKM})_{ij}\right)\,. \nonumber
\end{align}
We note that in general, $g_{H}$ is a complex parameter: for example, it is complex in LR models, see Eq.~\eqref{LRmodelcorrespondence}. However, the transformation $g_{H}\to g_{H}e^{-i\alpha}$ (with $\alpha$ an arbitrary phase) on the Lagrangian \eqref{lagrangian} only results, after diagonalization of $W$-$W'$ mixing, in $V_{R}\to e^{i\alpha}V_{R}$, therefore its effects are negligible for our scopes. Thus for simplicity we take $g_{H}$ to be real. The charged current interactions for leptons have the form
\begin{equation} 
\mathcal{L}^{\ell}_{cc}\,=\,W^{+}_{\mu}\cos\hat\theta\frac{g}{\sqrt{2}}\overline{\nu}^{i}_{L}\gamma^{\mu}e^{i}_{L}-W^{\prime\,+}_{\mu}\sin\hat\theta \frac{g}{\sqrt{2}}\overline{\nu}^{i}_{L}\gamma^{\mu}e^{i}_{L}\,.
\end{equation}
The trilinear couplings involving the $W'$, the $W$ and the Higgs and the $W'$ and two SM gauge bosons read
\begin{subequations}
\begin{align} \nonumber
\mathcal{L}_{W'Wh}&=\Big[-\frac{1}{2}g^{2}vh\sin\hat\theta\cos\hat\theta+\frac{g_{H}g}{\sqrt{2}}vh(\cos^{2}\hat\theta -\sin^{2}\hat\theta)+\frac{g_{4}^{2}}{2}hv\sin\hat\theta\cos\hat\theta\, \Big] \\
&\hspace{4mm}\times (W^{+\,\mu}W_{\mu}^{\prime\,-}+W^{-\,\mu}W_{\mu}^{\prime\,+})\,, \\
\label{W'Wgamma-111}
\mathcal{L}_{W'W\gamma}&=-i\,e(c_{B}+1)\sin\hat\theta \cos\hat\theta F_{\mu\nu}(W^{+\,\mu}W^{\prime\,-\,\nu}+W^{\prime\,+\,\mu}W^{-\,\nu})\,, \\
\mathcal{L}_{W'WZ}&=\, i\sin\hat\theta  \cos\hat\theta\Big[(g\cos\theta_{w}+g'\sin\theta_{w})(W^{-\,\mu}W^{\prime\,+}_{\nu\mu}+W^{\prime\, -\, \mu}W^{+}_{\nu\mu}- W^{\prime\,+\, \mu}W^{-}_{\nu\mu} \nonumber\\ \label{W'WZvertex}
&\hspace{4mm} -W^{+\,\mu} W^{\prime\,-}_{\nu\mu})Z^{\nu} -(g\cos\theta_{w}-g'\sin\theta_{w}c_{B})\left(W^{+\,\mu} W^{\prime\,-\,\nu}+ W^{\prime\,+\,\mu} W^{-\,\nu}\right) Z_{\mu\nu}\Big]\,,  
\end{align}
\end{subequations}
where $\theta_{w}$ is the weak mixing angle. Partial widths for decays into two-body final states are collected in App.~\ref{Partialdecaywidths}.

In summary, in addition to the $W'$ mass, 4 couplings appear in our phenomenological Lagrangian: $g_{q}$, $g_{H}$ (or equivalently the mixing angle $\hat\theta$), $g_{B}$ and $g_{4}$. We find it useful to normalize $g_{B}$ to the SM hypercharge coupling, so we will refer to $c_{B}\equiv g_{B}/g'$ in what follows. Our phenomenological Lagrangian describes the low energy limit of a LR model\footnote{Here we are assuming the $Z'$ to be sufficiently heavier than the $W'$, and we are neglecting effects coming from a different scalar spectrum.} for the following values of the parameters (see App.~\ref{minimalgaugemodels}):
\begin{equation}
\begin{array}{llll} \label{LRmodelcorrespondence}
g=g_{L}\,, \qquad &\displaystyle g^{\prime}=\frac{g_{X}g_{R}}{\sqrt{g_{X}^{2}+g_{R}^{2}}}\,, \qquad &g_{q}=g_{R}\,, \qquad &\displaystyle g_{H}=-2\sqrt{2}g_{R}\frac{kk'e^{-i\alpha_{1}}}{v^{2}}\,,\\\\
c_{B}=-1\,,\qquad &\displaystyle g_{4}^{2}=2g_{R}^{2}\frac{k^{2}+k^{\prime\,2}}{v^{2}}\,, \qquad &\displaystyle \tilde{M}^{2}=\frac{g_{R}^{2}v_{R}^{2}}{4}\,,\qquad  &v^{2}=2(k^{2}+k^{\prime\,2})\,.  
\end{array}
\end{equation}

\section{Indirect bounds} \label{indirect bounds}

 In this section we discuss indirect bounds on the couplings: $g_{q}$ is mainly constrained by $K$ and $B$ meson mixings, i.e. $\Delta F=2$ transitions (as we discuss in the next paragraph, the bounds are however strongly dependent on the structure of the right-handed mixing matrix $V_{R}$), while $\hat\theta$ is constrained by EWPT and by $u\rightarrow d$ and $u\rightarrow s$ transitions. The coupling $g_{B}$ is weakly constrained by Trilinear Gauge Couplings (TGC) measured at LEP, while $g_{4}$ is essentially unconstrained and marginal in our analysis (it only affects, and in a subleading way, the partial width for the decay $W'\rightarrow Wh$).  

\subsection{Bounds on the coupling to quarks $g_{q}$ from $\Delta F=2$ processes}

The heavy charged vector we are considering, being coupled to right-handed quark currents, contributes in general to the $K_{L}$-$K_{S}$ mass difference via box diagrams. The experimental determination of $\Delta m_{K}$ thus gives a constraint on the mass $M_{W'}$ and on the coupling of the $W'$ to quarks $g_{q}$; however, the bound has evidently a strong dependence on the assumed form for the right-handed quark mixing matrix $V_{R}$ (we remark that $V_{R}$ does not need to be unitary in our effective approach). It was shown in Ref.~\cite{Langacker:1989p2578} that for some special choices of $V_{R}$ the constraint is weakened significantly (notice that the discussion of Ref.~\cite{Langacker:1989p2578} was performed in the context of LR models, so unitarity of $V_{R}$ was assumed). We choose for our phenomenological analysis the least constrained of these special forms, namely 
\begin{equation} \label{RHmixing}
\left|V_{R}\right|=\mathbb{1}\,,
\end{equation}
for which the bound reads at $90 \%$ CL \cite{Langacker:1989p2578}
\begin{equation} \label{Klimit}
M_{W'}> \frac{g_{q}}{g}\,300\,\mathrm{GeV}\,.
\end{equation} 
We note that in specific models, the bound can be much stronger: for example, if a discrete symmetry (P or C) relating the left and right sectors is imposed in LR models, then the bound reads approximately $M_{W'}>(2\,\text{--}\,3)\,\textrm{TeV}$ (see, e.g., Refs.~\cite{Zhang:2007da,Maiezza:2010ic}). This happens because the discrete symmetry forces $V_{R}$ to be close to $V_{CKM}$, implying a mixing of the order of the Cabibbo angle between the first two generations. 

While mixing among the first two families is strongly constrained by $K_{L}$-$K_{S}$ data, we could consider the case where significant mixing between the first and third, or between the second and third families is present; we should accordingly take into account the constraints from $B$ meson physics, as discussed in Ref.~\cite{Frank:2010p3284}, where constraints on the elements of the right-handed mixing matrix from $b\rightarrow s\gamma$ and from $B^{0}_{d,s}$-$\overline{B}^{0}_{d,s}$ mixing were analysed in the context of a LR model. However, this goes beyond the scope of our work, so we simply take the form \eqref{RHmixing}, which automatically satisfies the constraints from $B$ meson mixing. The corresponding upper bound from $K$ mixing, Eq.~\eqref{Klimit}, is negligible with respect to the constraints coming from Tevatron direct searches (see Section~\ref{Tevatron}). Also notice that, as discussed in Ref.~\cite{Langacker:1989p2578}, this bound still holds if each $(V_{R})_{ij}$ is varied of $\epsilon=0.01$ from its central value, so that extreme fine tuning is avoided. For a study of the LHC phenomenology of a LR model with large off-diagonal $V_{R}$ elements, see Ref.~\cite{Frank:2010p2250}.   

\subsubsection{Flavor-violating $W'$ as an explanation of the top $A_{FB}$ puzzle}
The `anomaly' observed by CDF in the forward-backward asymmetry of top pairs has recently drawn a lot of attention. The most recent measurement of $A^{t\overline{t}}_{FB}$ found a discrepancy of around $2\sigma$ with respect to the SM prediction  \cite{CDFCollaboration:2010p3438}; furthermore, the asymmetry is observed to be larger in the region of large invariant mass of the $t\overline{t}$ pair, and in the region of large rapidity difference $|y_{t}-y_{\overline{t}}|$. The $t$-channel exchange of a $W'$ that only couples to $t$ and $d$ quarks was suggested in Ref.~\cite{Jung:2010p015004} as a possible explanation of the anomaly, and in Refs.~\cite{Cheung:2009p3324,Cheung:2011qa} it was shown that the observed asymmetry can be reproduced with the introduction of a right-handed $W'$ with mass in the range $200\,\text{--}\, 600$ GeV, and coupling $W'$-$t$-$d$ of magnitude $0.85\,\text{--}\, 2.1$. Similar values were chosen in Refs.~\cite{Shelton:2011p3339,Barger:2011ih}. Such $W'$ is described by our framework, where the right-handed mixing matrix does not need to be unitary, and as a consequence can accommodate a large $W'$-$t$-$d$ coupling, while having the remaining entries tuned to evade, e.g., the strong bounds coming from meson mixing.

\subsection{Bounds on the $W$-$W'$ mixing angle $\hat{\theta}$}

The main constraints on the $W$-$W'$ mixing angle $\hat\theta$ come from EWPT and from semileptonic $u\rightarrow d$ and $u\rightarrow s$ transitions. The $W$-$W'$ mixing term in \eqref{V-SM} breaks custodial symmetry, and is therefore strongly constrained by EWPT. A recent electroweak fit (including LEP2 data) performed in Ref.~\cite{Aguila:2010p1781} gives at $95\%$ CL
\begin{equation} \label{LEPbound}
\left|\frac{g_{H}}{M}\right|< 0.11\, \mathrm{TeV}^{-1}\,.
\end{equation}
We have checked that, as already remarked in Ref.~\cite{Aguila:2010p1781}, this constraint is essentially due to the negative contribution the $W$-$W'$ mixing gives to the $T$ parameter: the leading term in the $v^{2}/M^{2}$ expansion reads
\begin{equation} \label{Tcontribution}
\hat{T}_{V}=-\frac{\Delta^{4}}{M^{2}m_{\hat{W}}^{2}}\,.
\end{equation}
The LEP2 lower limit on the Higgs mass thus forces such a contribution to be very small. The bound \eqref{LEPbound} was in fact computed in Ref.~\cite{Aguila:2010p1781} leaving the Higgs mass as a free fit parameter, and including data from direct Higgs searches at LEP2. The results of our study depend very weakly on the mass of the Higgs, as long as it is light. We can translate Eq.~\eqref{LEPbound} into an upper bound on $\hat\theta$: the resulting limit becomes stronger when the mass of the $W'$ is increased, and varies from $|\hat\theta|\lesssim 4\times 10^{-3}$ for $M_{W'}=300\,\text{GeV}$ to $|\hat\theta|\lesssim 5\times 10^{-4}$ for $M_{W'}=2$ TeV.

A bound on the mixing angle $\hat\theta$ of different origin comes from the precise \mbox{low-energy} measurement of $u\rightarrow d$ and $u\rightarrow s$ transitions (i.e. from the measurements of the corresponding entries of the CKM matrix). Integrating out both the $W$ and the $W'$, we obtain a four-fermion effective Lagrangian that can be used to compute constraints from such measurements. The operators relevant to semileptonic processes, which give the strongest bounds, are
\begin{equation}
\mathcal{L}_{\text{eff}}=-\frac{4G_{F}}{\sqrt{2}}\,\overline{u}\,\gamma^{\mu}\Big[(1+\epsilon_{L})V_{CKM}P_{L}+\epsilon_{R}V_{R}P_{R}\Big]d\,(\overline{\ell}_{L}\gamma_{\mu}\nu^{\ell}_{L})+\text{h.c.}\,,
\end{equation}
where, neglecting $O(v^{4}/M^{4}_{W'})$ terms, $\epsilon_{L}=0$ and $\epsilon_{R}=g_{q}\,\hat\theta/g$. In Ref.~\cite{Buras:2010p3239} the bound $\epsilon_{R}\,\mathrm{Re}(V_{R}^{ud})=(0.1\pm 1.3)\times 10^{-3}$ was obtained, which assuming small CP phases implies at 95$\%$ CL
\begin{equation}
-2\times 10^{-3} < \epsilon_{R}V_{R}^{ud} < 3\times 10^{-3}\,.
\end{equation}
On the other hand, such bound is strongly relaxed if CP phases in $V_{R}$ are large: in the limit of maximal CP phases, only a milder second-order constraint survives, leading (assuming $V_{R}^{ud}\approx 1$) roughly to $|\epsilon_{R}|<10^{-(2\,\text{--}\,1)}$, as discussed in Ref.~\cite{Langacker:1989p2578}. 

By making use of soft-pion theorems, constraints from nonleptonic processes such as $K\to 2\pi$ and $K\to 3\pi$ were also computed \cite{Donoghue:1982mx,Bigi:1981rr}. However, such bounds were obtained neglecting long distance chiral loop effects, which are known to be important and can offset tree-level results. Therefore, we do not consider such constraints in the following.

\subsection{Bounds from trilinear gauge couplings}
The $WWV_{0}$ vertex ($V_{0}=\gamma, Z$) can be described, assuming C- and P- conservation, by an effective Lagrangian containing 6 parameters (see for example Ref.~\cite{TripleGaugeCouplongsWorkingGroup:1996p2576}):
\begin{equation*}
\mathcal{L}_{\text{eff}}^{WWV_{0}}=ig_{WWV_{0}}\Big[g_{1}^{V_{0}}V_{0}^{\mu}(W^{-}_{\mu\nu}W^{+\nu}-W^{+}_{\mu\nu}W^{-\nu})+k_{V_{0}}W^{+}_{\mu}W_{\nu}^{-}V_{0}^{\mu\nu}+\frac{\lambda_{V_{0}}}{m^{2}_{W}}V_{0}^{\mu\nu}W_{\nu}^{+\rho}W^{-}_{\rho\mu}\Big]
\end{equation*}
where $g_{WW\gamma}=e,\,g_{WWZ}=g\cos\theta_{w}$, and the SM values of the parameters are given by $g_{1}^{\gamma,Z}=\kappa_{\gamma,Z}=1$ and $\lambda_{\gamma,Z}=0$. Assuming $SU(2)_{L}\times U(1)_{Y}$ gauge invariance reduces the number of independent parameters to three, which can be taken to be $\Delta g_{1}^{Z}\equiv g_{1}^{Z}-1$, $\Delta k_{\gamma}\equiv k_{\gamma}-1$, and $\lambda_{\gamma}$. In the case under discussion, the expressions of these parameters read
\begin{equation}
\Delta g_{1}^{Z}=-\sin^{2}\hat\theta(1+\tan^{2}\theta_{w})\,,\qquad \Delta k_{\gamma}=-\sin^{2}\hat\theta(1+c_{B})\,,\qquad \lambda_{\gamma}=0\,.
\end{equation}
Thus we can use the fits to LEP2 data performed by the LEP experiments \cite{DELPHICollaboration:2010p2577,ALEPHCollaboration:2004p0726,L3Collaboration:2002p151,L3Collaboration:2004p151,OPALCollaboration:2004p463} letting $\Delta g_{1}^{Z},\Delta k_{\gamma}$ free to vary while keeping fixed $\lambda_{\gamma}=0$, to constrain the values of our model parameters $(c_{B},\hat\theta)$. By combining this limit with the upper bound on the mixing angle $\hat\theta$ presented in the previous subsection, we can in principle constrain $c_{B}$. However, since as discussed above the mixing angle is required to be very small, in practice TGC constrain only extremely weakly the value of $c_{B}$. For example, using the analysis performed by the DELPHI Collaboration \cite{DELPHICollaboration:2010p2577}, we find that even considering a very large mixing angle $|\hat\theta| \sim 10^{-1}$, the wide range $-11 < c_{B}< 20$ (i.e. $-3.9< g_{B}< 7.1$) is allowed by TGC measurements at $95\%$ CL.

\section{Bounds from Tevatron direct searches} \label{Tevatron}

\begin{figure}[t]
\includegraphics[scale=0.34]{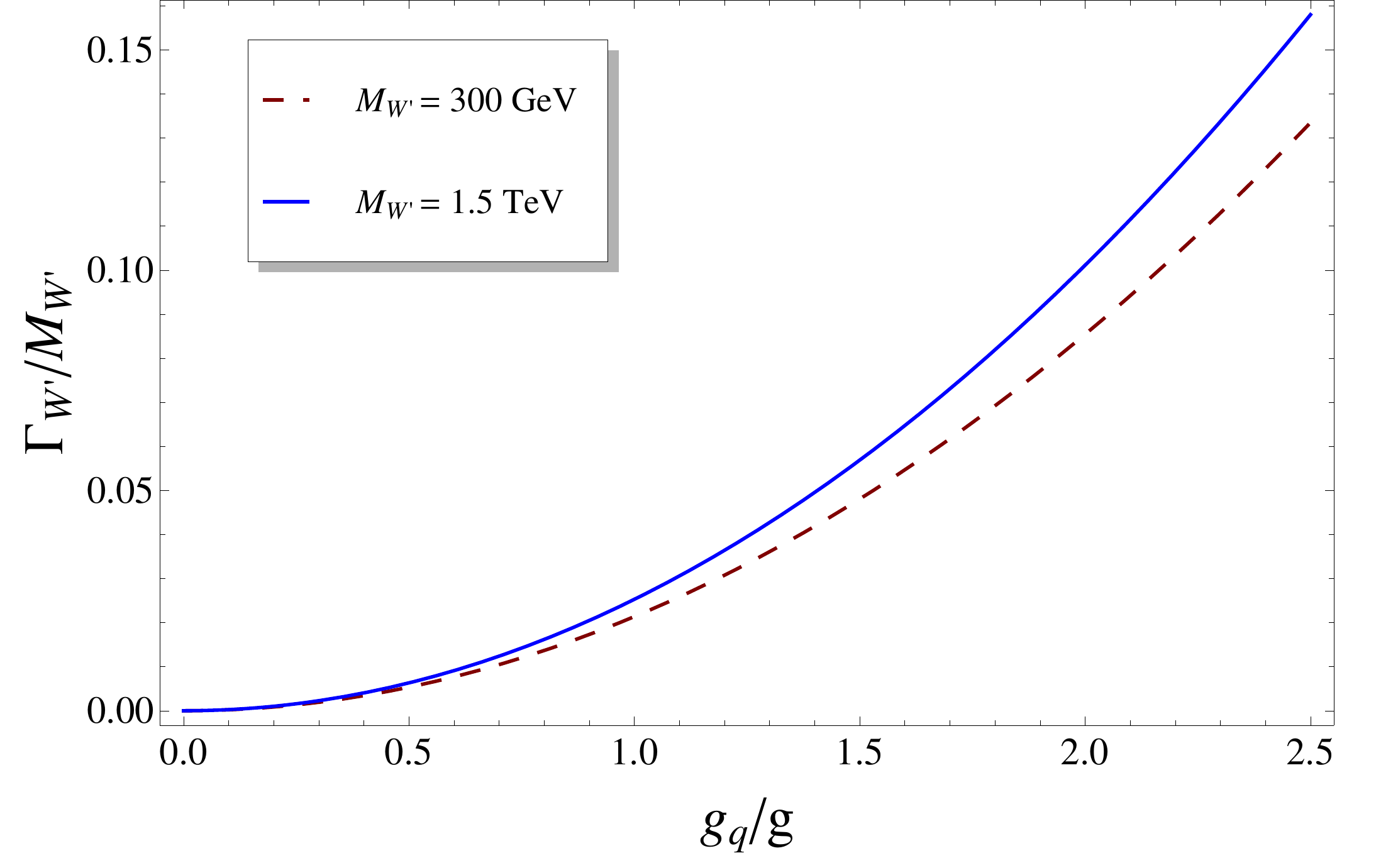}
\hspace{1mm}\includegraphics[scale=0.34]{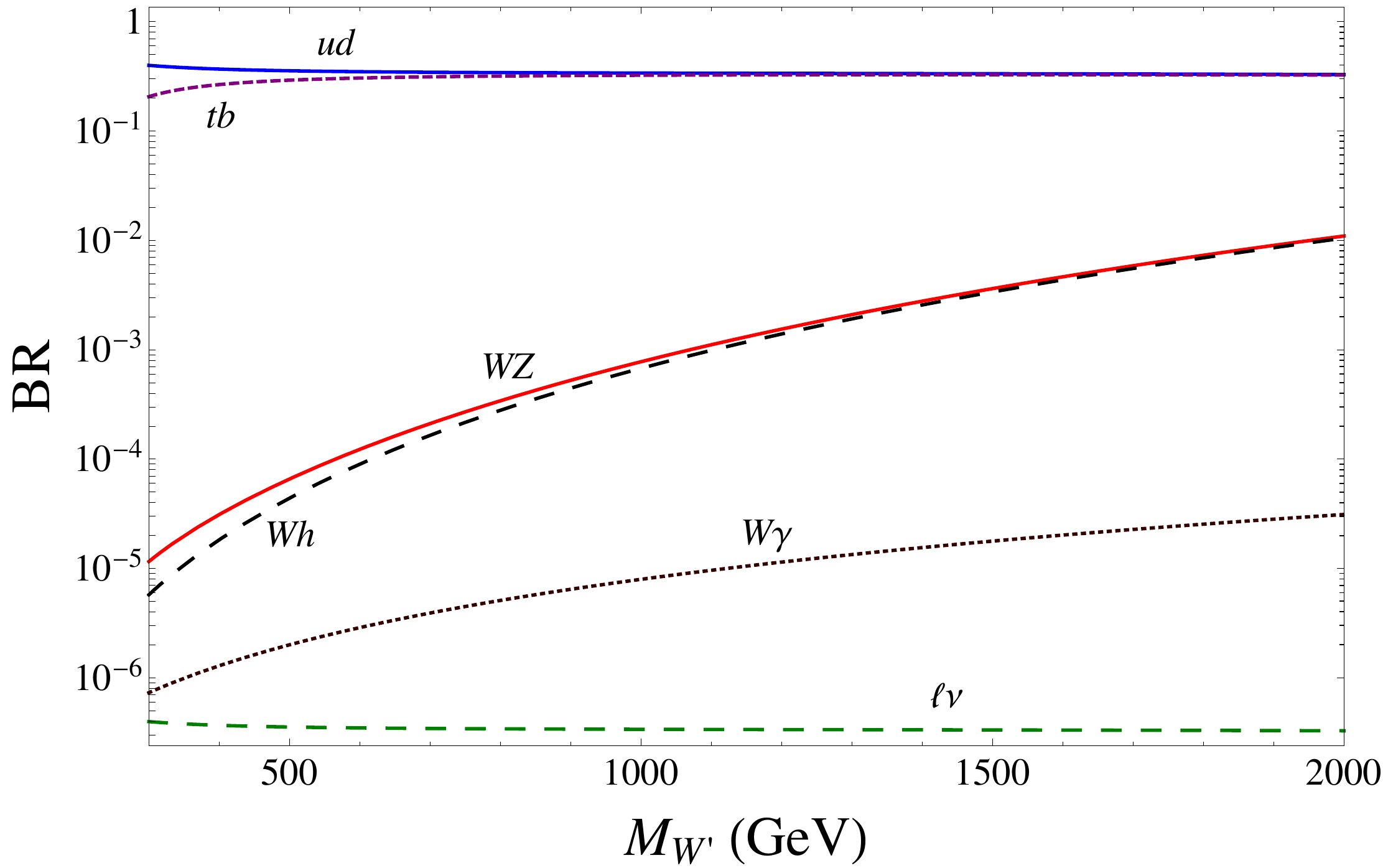}
\caption{\textsl{Left panel.} $W'$ width over mass ratio as a function of $g_{q}/g$  for negligible mixing, $\hat\theta \approx 0$, for $M_{W'}=300\,\mathrm{GeV}$ (dashed, red) and 1.5 TeV (blue). \textsl{Right panel.} Branching ratios of the $W'$ as a function of its mass, for the following choice of the remaining parameters: $g_{q}=g$, $\hat\theta=10^{-3}$, $c_{B}=-3$, $g_{4}=g$. From top to bottom: $ud$, $tb$, $WZ$, $Wh$, $W\gamma$, $\ell\nu$ (the latter includes all the three lepton families).}
\label{fig:totalwidth}
\end{figure}

Data collected by the CDF and D0 experiments at the Tevatron in the dijet and $tb$ final states\footnote{By $tb$ we will always mean the sum $t\overline{b}+\overline{t}b$.} can be used to set an upper limit on the coupling to quarks of the $W'$ we are discussing as a function of its mass. In this section, we assume negligible $W$-$W'$ mixing, $\hat\theta \approx 0$, so the only relevant parameters are the $W'$ mass and the coupling $g_{q}$, and we obtain an upper bound on $g_{q}$ as a function of $M_{W'}$. If $W$-$W'$ mixing happens to be sizable, then the branching ratio into quarks is reduced, and the upper bound on $g_{q}$ gets relaxed accordingly. For instance, taking the relatively large value $\hat\theta = 10^{-2}$, the upper bound on $g_{q}$ is relaxed by approximately $10\%$ for $M_{W'}\gtrsim 1\,\mathrm{TeV}$, and less for lighter $W'$. The dependence of the ratio $\Gamma_{W'}/M_{W'}$ on the coupling $g_{q}$ is plotted in the left panel of Fig.~\ref{fig:totalwidth}, while the branching ratios as functions of $M_{W'}$ are shown in the right panel of the same figure, for representative values of the parameters.

Unless explicitly noted, we take the latest average value of the top mass, namely $m_{t}=173.3$ GeV \cite{TevatronElectroweakWorkingGroup:2010p3432}, and make use of the CTEQ6L set of parton distribution functions \cite{Pumplin:2002p3433}. Cross sections are computed using the CalcHEP matrix element generator \cite{Pukhov:1999p1261,Calchepweb}.
\subsection{Dijet final state}
\begin{figure}[t]
\includegraphics[scale=0.34]{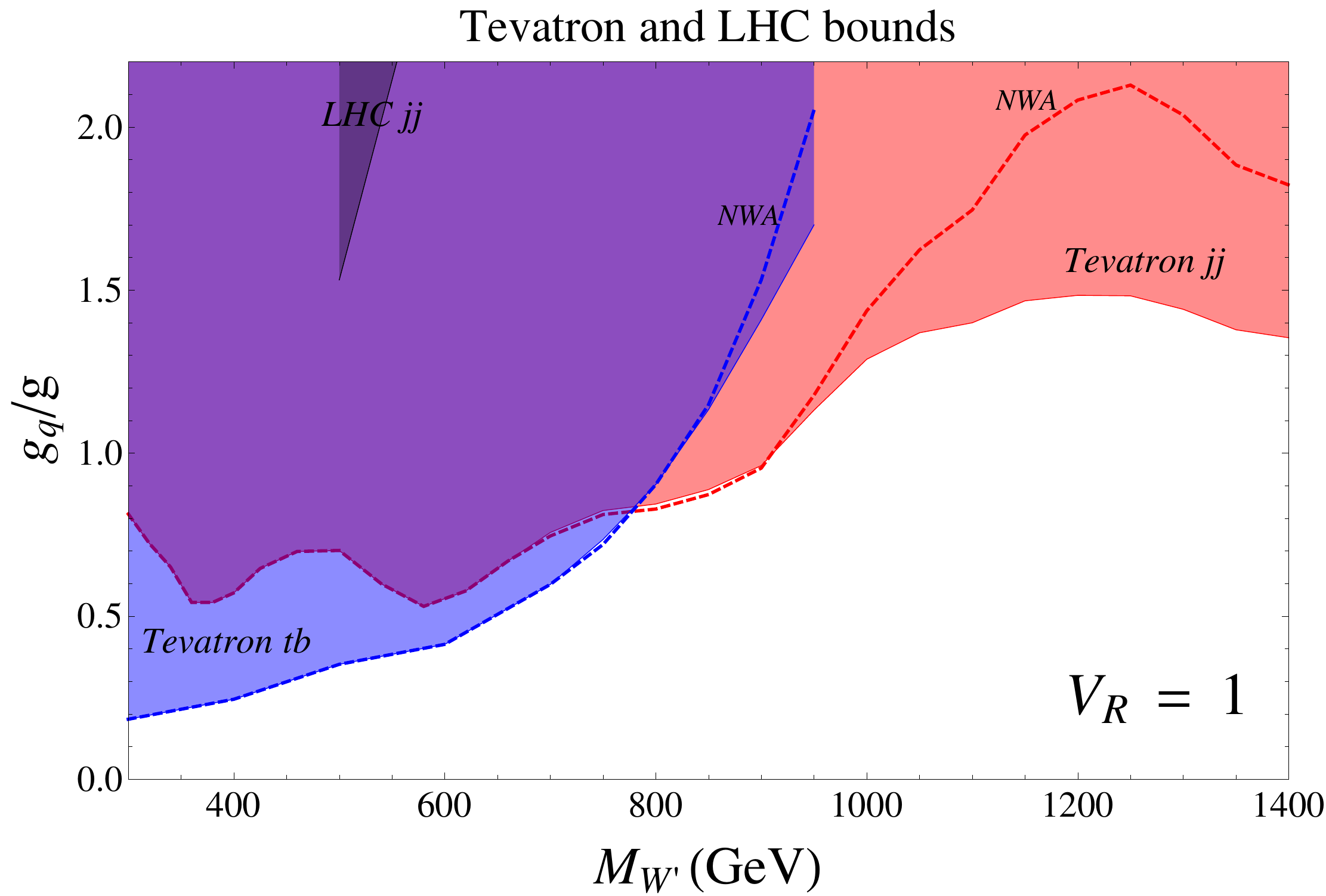}
\hspace{2.5mm}\includegraphics[scale=0.34]{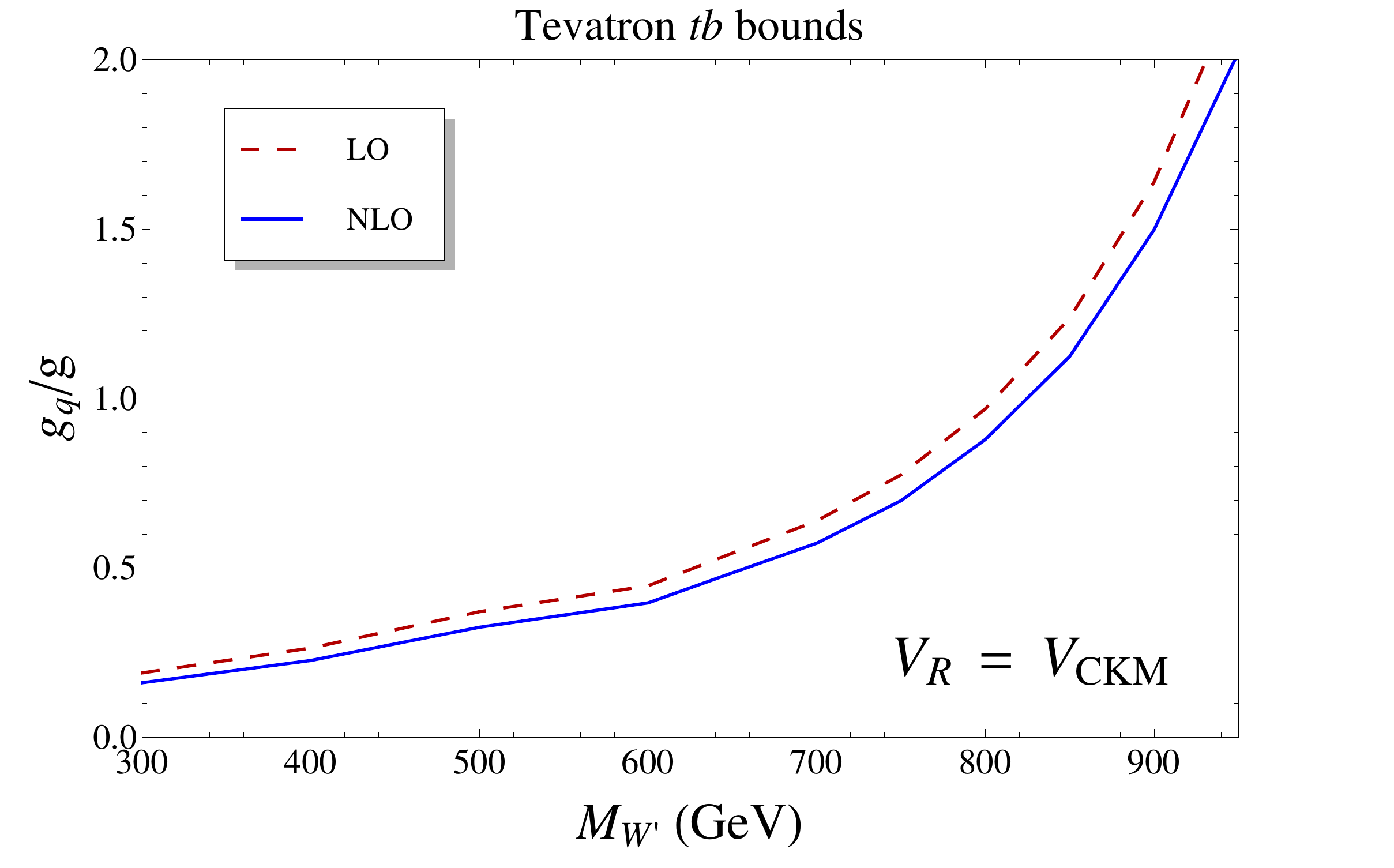}
\caption{\textit{Left panel}. The region of the ($M_{W'},g_{q}/g$) plane excluded at $95\%$ CL by Tevatron searches in the dijet final state (red region extending up to $1.4\,\mathrm{TeV}$) and $tb$ final state (blue region extending up to $950\,\mathrm{GeV}$). The dashed lines correspond to exclusion limits computed assuming $\sigma\propto g_{q}^{2}$, see text for details. Also shown in grey is the region excluded at $95\%$ CL by CMS dijet searches, see Section~\ref{LHC}. \textit{Right panel}. Upper limit from $tb$ searches on the $W'$ coupling to quarks as a function of $M_{W'}$, obtained using cross sections at LO (dashed) and at NLO (continuous) as reported in Ref.~\cite{Sullivan:2002p2617}. The scaling behaviour $\sigma\propto g_{q}^{2}$ was assumed.}
\label{fig:Tevatronlimits}
\end{figure}

Searches for resonances in the invariant mass spectrum of dijet events at CDF and D0 are sensitive to the $W'$ we are discussing, which decays into quarks with branching ratio close to unity. The most recent dijet search, based on 1.13 fb$^{-1}$ of data, has been performed by the CDF collaboration \cite{CDFCollaboration:2009p2710}. Since no discrepancy with the SM prediction was observed, upper limits on the product $\sigma(p\overline{p}\rightarrow W'\rightarrow jj)\times \mathcal{A}$, where $\mathcal{A}$ is the geometrical acceptance for having both jets with $|y|<1$, have been set in Ref.~\cite{CDFCollaboration:2009p2710} for several types of resonance, including a $W'$. Therefore, we can compute $\sigma(p\overline{p}\rightarrow W'\rightarrow jj)\times \mathcal{A}$ using our phenomenological Lagrangian, and extract an upper bound on $g_{q}$ for each value of $M_{W'}$, which is reported in the left panel of Fig.~\ref{fig:Tevatronlimits}. We use cross sections at Leading Order (LO).
The acceptance $\mathcal{A}$ is $36\%$ at $M_{W'}=300\, \mathrm{GeV}$, reaches a maximum of $51\%$ for $M_{W'}\sim 800$ GeV, and decreases for larger masses, being $34\%$ at $1.4$ TeV. The decreasing behavior of the acceptance at high resonance masses is due to a threshold effect: for a $W'$ mass around $1\, \mathrm{TeV}$ and above (that is, close to the kinematic limit of the Tevatron, which has a center of mass energy of $1.96\,\mathrm{TeV}$), the probability that the on-shell production condition $x_{1}x_{2}\approx M_{W'}^{2}/s$ is satisfied is so small that the off-shell contribution to the $p\overline{p}\rightarrow W'\rightarrow jj$ cross section becomes relevant, making the acceptance behave differently from what we would naively expect for an on-shell production mechanism. To make the relevance of this threshold effect more manifest, we also plot the upper bound on the coupling obtained by rescaling, for each value of $M_{W'}$, the value of $\sigma\times \mathcal{A}$ computed for $g_{q}=g$ according to the relation $\sigma\times \mathcal{A}\propto g_{q}^{2}\,$, which holds exactly in the Narrow Width Approximation (NWA). It is evident that while for masses below approximately 800 GeV the NWA (pure \mbox{on-shell} production) provides an excellent description of the dijet resonant production, for a larger mass of the resonance the NWA is not reliable anymore. Because of these relevant off-shell effects, we prefer to avoid using the notation $\sigma(W')\times \mathrm{BR}(W'\rightarrow jj)$. 

The method we use to compute limits is valid for a resonance width smaller than the dijet energy resolution, which for the CDF experiment is of the order of $10\%$ of the dijet mass. The $W'$ we are studying has a width of $\sim 10\%$ of its mass for $g_{q}\sim2g$, as can be read off the left panel in Fig.~\ref{fig:totalwidth}; for larger couplings, the resonance width cannot be neglected, and the analysis would need to be corrected for this effect. 

\subsection{$tb$ final state}
Another final state which is relevant to our model is $tb$. The CDF and D0 collaborations have searched for narrow resonances decaying into $tb$, with the $W$ coming from the top decaying into a lepton and missing transverse energy. The most recent search from CDF is based on 1.9 fb$^{-1}$ of data \cite{CDFCollaboration:2009p3125}, whereas D0 has carried out a similar analysis with 0.9 fb$^{-1}$ \cite{D0Collaboration:2008p2709}\footnote{The latest D0 analysis is based on 2.3 fb${}^{-1}$ of data \cite{Abazov:2011xs}. However, unfolding of the cross section limits, which is necessary to interpret them in our framework, was performed by the D0 collaboration only after completion of this work. Therefore the bounds from Ref.~\cite{Abazov:2011xs} are not included here. We expect that they will be slightly more stringent than those presented in Fig.~\ref{fig:Tevatronlimits}.}. Both analyses give as result upper limits on $\sigma(p\overline{p}\rightarrow W'\rightarrow tb)$, so we can compute the latter quantity using our phenomenological Lagrangian to extract an upper bound on $g_{q}$ for each value of the $W'$ mass. The strongest constraints are given by the CDF analysis, and in the left panel of Fig.~\ref{fig:Tevatronlimits} we compare them with the dijet limits discussed in the previous subsection. Analogously to what happened for the dijet final state, for $M_{W'}\gtrsim 800$ GeV threshold effects become relevant, and correspondingly the upper limit on $g_{q}$ computed assuming the NWA relation $\sigma\propto g_{q}^{2}$ differs from the correct limit. 

\subsubsection{Comparison of LO and NLO $tb$ limits}
We have carried out our analysis at LO. However, for illustration purposes, it is useful to compare in Fig.~\ref{fig:Tevatronlimits} the upper bound on $g_{q}$ from $tb$ searches computed using cross sections at NLO and at LO, both as given in Ref.~\cite{Sullivan:2002p2617}. Since in the latter paper all cross sections were computed for $g_{q}=g$, we assume the relation $\sigma(p\overline{p}\rightarrow W'\rightarrow tb)\propto g_{q}^{2}$, which is exact in the NWA\footnote{We stress, however, that deviations from the NWA, and as a consequence from the $\sigma\propto g_{q}^{2}$ behaviour, arise due to off-shell effects for $M_{W'} \gtrsim 800$ GeV, as already discussed in the previous paragraph.}, to extract the upper bound on the coupling. Notice that the dependence on $M_{W'}$ of the NLO upper bound on $g_{q}$ differs significantly from those reported by the CDF and D0 collaborations in Refs.~\cite{CDFCollaboration:2009p3125} and \cite{D0Collaboration:2008p2709} respectively: this is due to the wrong assumption made there, that the cross section is proportional to the fourth power of the coupling. We remark that the upper limits shown in the right panel of Fig.~\ref{fig:Tevatronlimits} are obtained assuming $m_{t}=175$ GeV and $V_{R}=V_{CKM}$ (see Ref.~\cite{Sullivan:2002p2617}); even though the difference is at the level of a few percent, for the sake of consistency in what follows we will use the limits computed with $m_{t}=173.3$ GeV and $V_{R}=\mathbb{1}$, and reported in the left panel of Fig.~\ref{fig:Tevatronlimits}.

\section{LHC phenomenology} \label{secLHC}

In this section we discuss the reach of the early LHC on the composite $W'$ we are studying. We analyse first the prospects for discovery of the resonance as an excess of events in the dijet invariant mass spectrum, and subsequently move on to discuss decays into two gauge bosons. We study first the $W'\rightarrow W\gamma$ decay, which is of special interest since it is strongly suppressed in gauge models. As a consequence, its observation would be a hint of the compositeness of the $W'$. Finally, we discuss the $W'\rightarrow WZ$ channel.   

\subsection{Dijet searches} \label{LHC}

The search for resonances in the dijet mass spectrum is one of the first new physics analyses performed by the CMS \cite{CMSCollaboration:2010p2595} and ATLAS \cite{ATLASCollaboration:2010p1746,ATLAS:2010Upd} experiments at the LHC, with an integrated luminosity of 2.9 and 3.1 pb${}^{-1}$ respectively at 7 TeV. Due to the very small data sample analysed so far, such searches are not competitive yet with those performed at the Tevatron: from Fig.~\ref{fig:Tevatronlimits} we see that only in a very narrow interval around $M_{W'}\sim 500$ GeV does the CMS search place a meaningful (even if weaker than the Tevatron one) upper limit on the $W'$ coupling to quarks. For larger masses, the CMS upper bound on the $W'$ cross section is saturated for values of the coupling $g_{q}>2g$, which implies that the width of the resonance is larger than the dijet mass resolution, and as a consequence the experimental analysis would need to be modified to account for a broad resonance. We use the CMS results because their limits were computed also for resonances decaying into a $qq$ final state, while the ATLAS analysis only assumes a resonance decaying into the final state $qg$, which leads to more radiation and as a consequence to a broader resonance shape, which has an effect on the cross section limits.

\begin{figure}[t]
\includegraphics[scale=0.34]{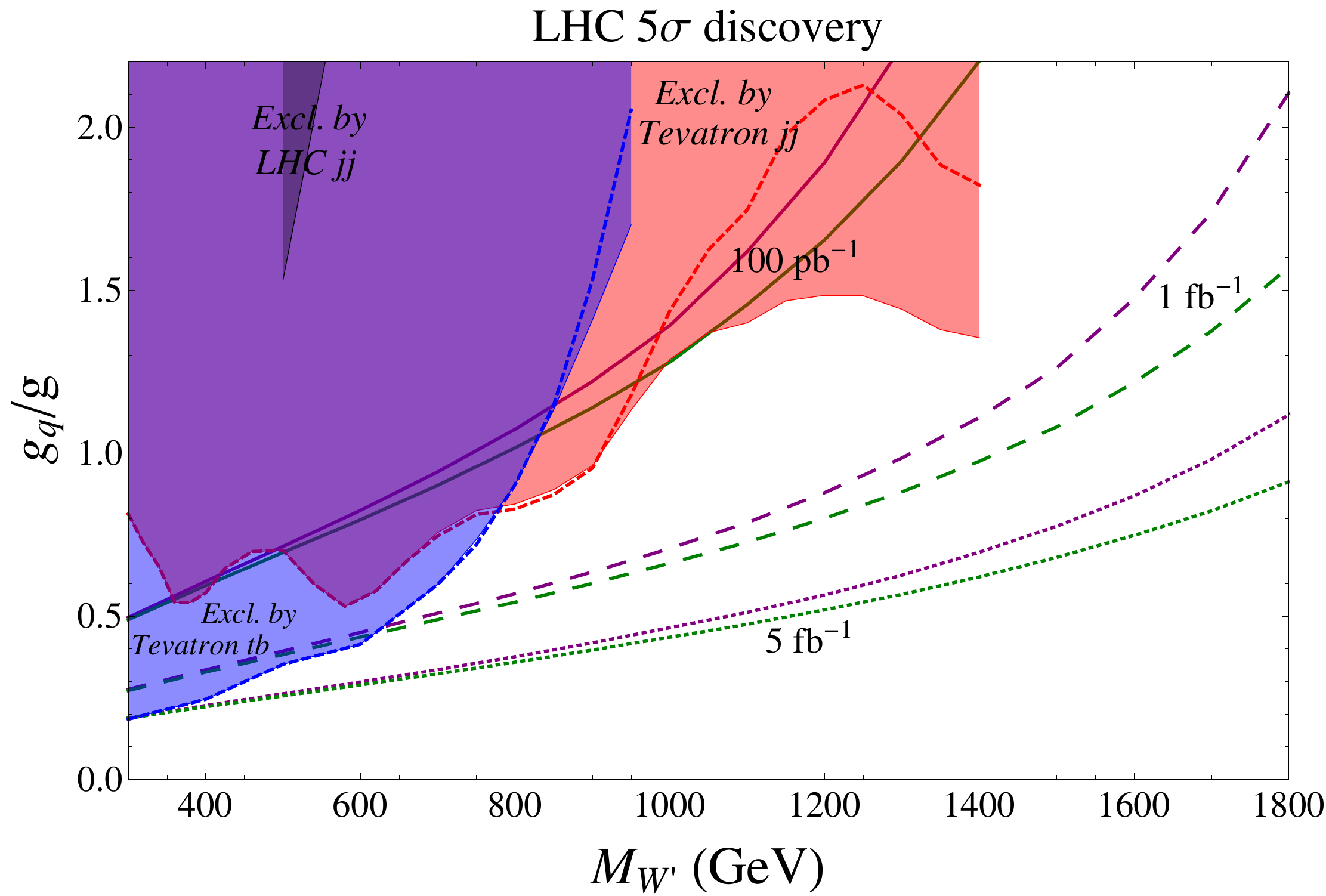}
\hspace{2mm}\includegraphics[scale=0.33]{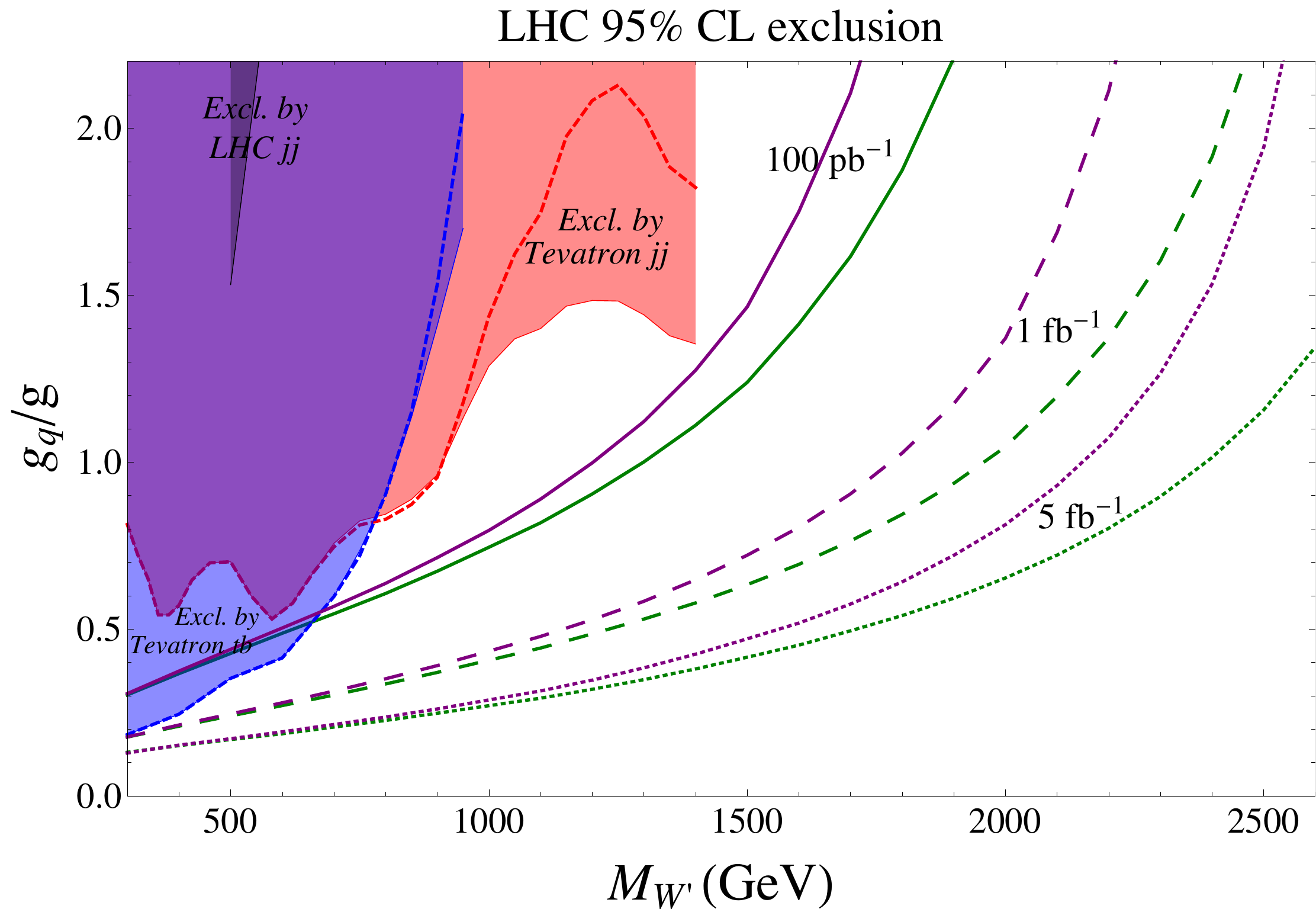}
\caption{Contours in the $(M_{W'},g_{q}/g)$ plane for $5\sigma$ discovery (left) and $95\%$ CL exclusion (right) at the 7 TeV and 8 TeV LHC, for an integrated luminosity of $L=0.1,\,1$ and $5$ fb${}^{-1}$, corresponding to the continuous, dashed and dotted lines, respectively (for each different dashing, the upper, purple line is for $7$ TeV and the lower, green line is for $8$ TeV).  Also shown are the Tevatron dijet (red) and $tb$ (blue) exclusions, together with the CMS exclusion with $2.9$ pb${}^{-1}$ (grey).}
\label{fig:LHCcouplingVSmass}
\end{figure}

\begin{figure}[t]
\includegraphics[scale=0.335]{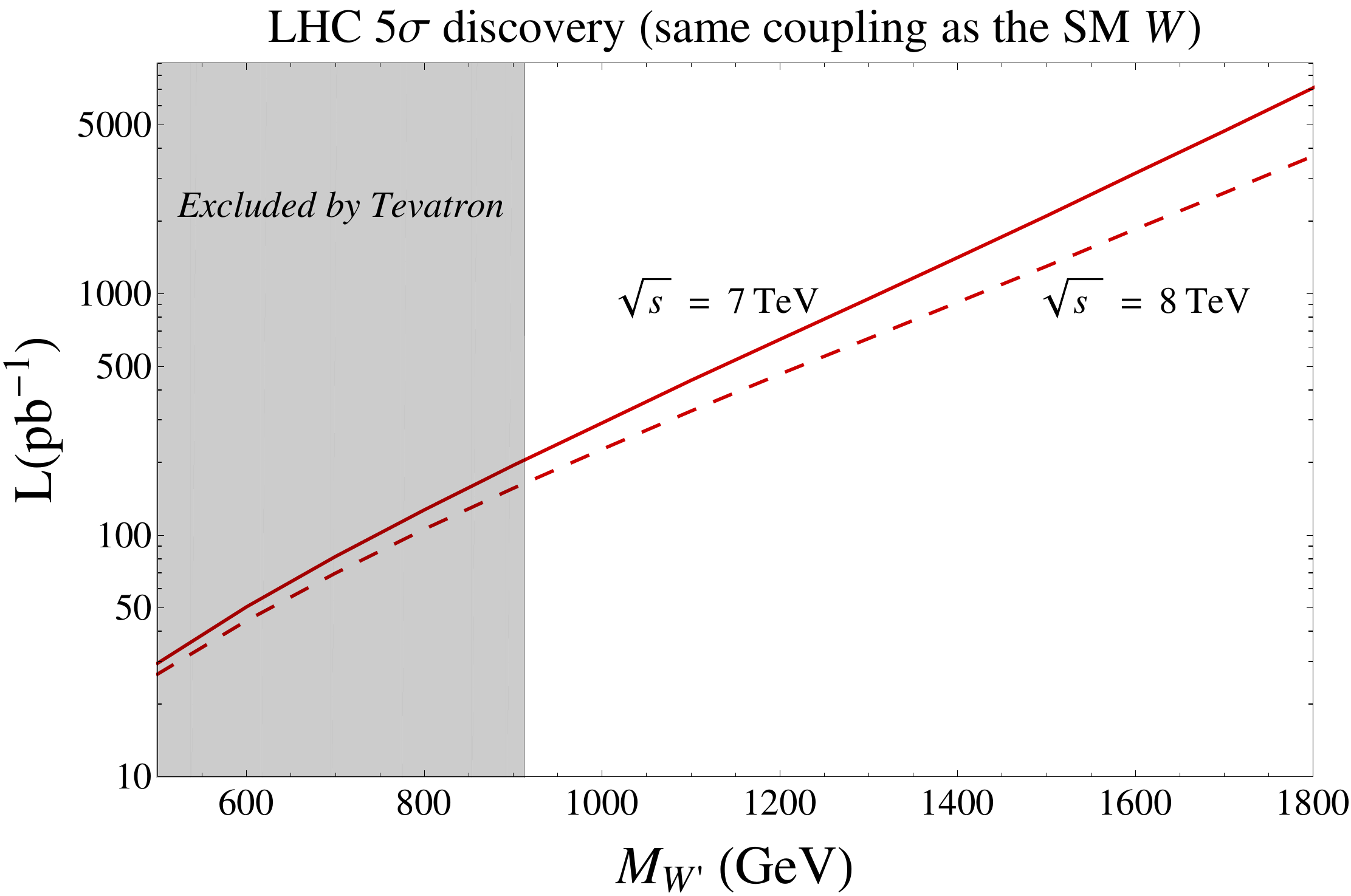}
\hspace{3mm}\includegraphics[scale=0.33]{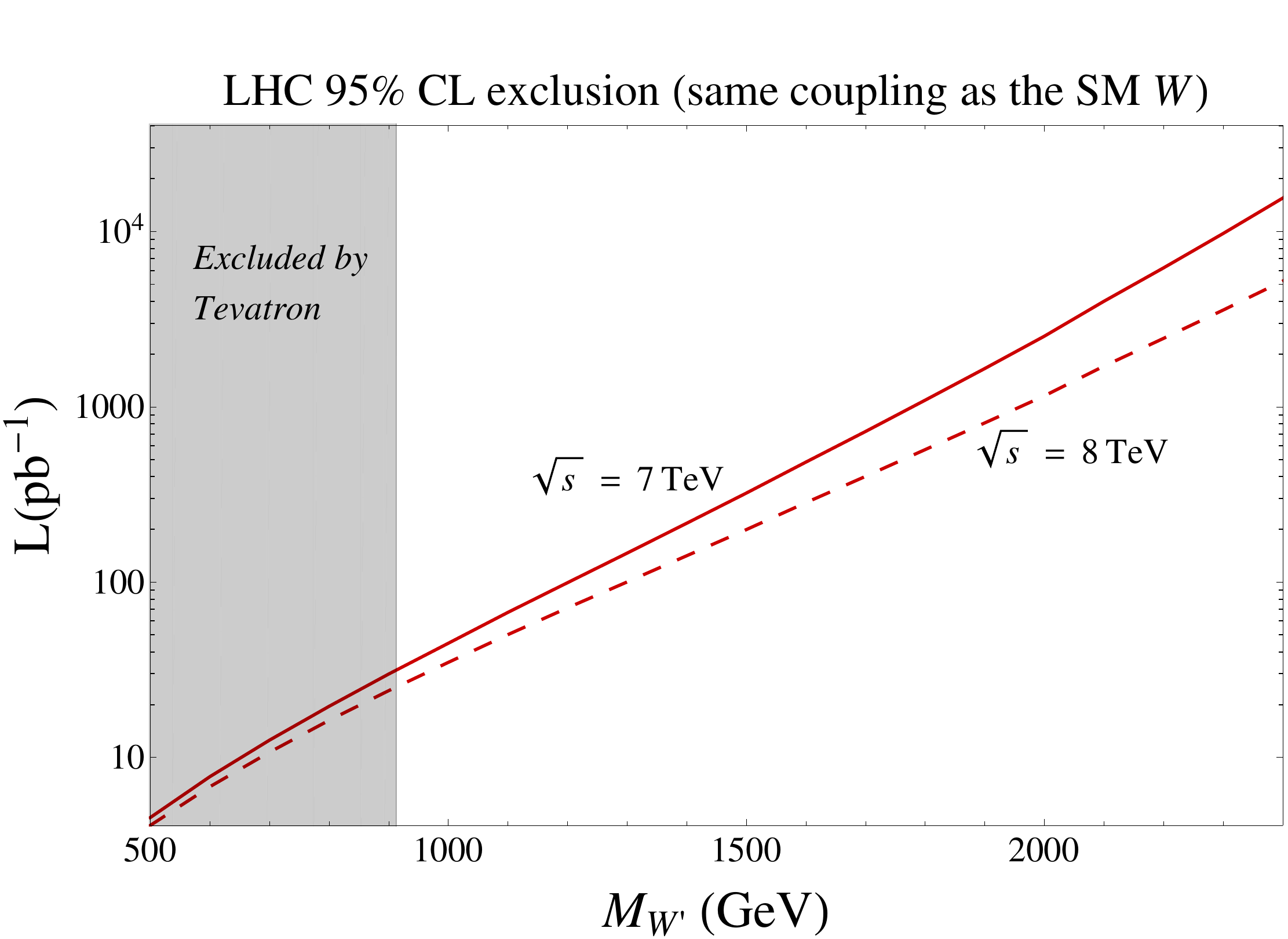}
\caption{Integrated luminosity needed for $5\sigma$ discovery (left) and $95\%$ CL exclusion (right) as a function of the $W'$ mass, for the 7 TeV (continuous) and 8 TeV (dashed) LHC. The region shaded in grey, corresponding to $M_{W'}<913$ GeV, is excluded at $95\%$ CL by Tevatron searches.}
\label{fig:LHCluminosityVSmass}
\end{figure}

Future LHC analyses, however, will soon overtake the Tevatron results, so it is interesting to discuss the reach of dijet searches on the $W'$ we are considering. We assume the CMS kinematic cuts, namely on the pseudorapidity $|\eta|<2.5$ of each jet, and on the pseudorapidity difference $|\Delta\eta|<1.3$ \cite{CMSCollaboration:2010p2595}. For values of $M_{W'}$ between \mbox{300 GeV} and 2.6 TeV, in intervals of 100 GeV, we compute as a function of the coupling $g_{q}$ the integral of the signal differential invariant mass distribution $d\sigma_{S}/dM_{jj}$ over the region $M_{jj}>M_{W'}(1-\epsilon/2)$, and compare the result with the integral of the background distribution over the same range, to obtain $5\sigma$ discovery and 95$\%\, \mathrm{CL}$ exclusion contours in the $(M_{W'},g_{q}/g)$ plane. Here $\epsilon$ is the dijet mass resolution, which following Ref.~\cite{CMSCollaboration:2010p2595} we assume to vary from 8$\%$ at $M_{W'}=500$ GeV to 5$\%$ at $2.5\,\mathrm{TeV}$. The results are shown in Fig.~\ref{fig:LHCcouplingVSmass} for three different integrated luminosities, namely \mbox{$L=\int\mathcal{L}=0.1,1,5$ fb${}^{-1}$}, and for two LHC center of mass energies, namely 7 and 8 TeV\footnote{It has recently been decided that the LHC will run at 7 TeV in 2011. However, a higher energy for 2012 cannot be excluded at the time of writing \cite{Carli:2010zz}.}. We find that \mbox{100 pb${}^{-1}$} are not sufficient for a discovery, even at 8 TeV (except perhaps for a very small region around $M_{W'}=1\,\mathrm{TeV}$). On the other hand, if we focus on the exclusion contours, we see that the LHC can do better than the Tevatron already with 100 pb${}^{-1}$ for \mbox{$M_{W'}\gtrsim 700\,\text{GeV}$}, and for essentially all $W'$ masses if the luminosity is increased to $1\,\text{fb}^{-1}$.
We also report in Fig.~\ref{fig:LHCluminosityVSmass}, as a function of $M_{W'}$, the integrated luminosity needed for discovery or exclusion of a $W'$ with coupling to quarks equal to that of the SM $W$($g_{q}=g$), both for the 7 and 8 TeV LHC.  
 
We choose to compare the integrals over $M_{jj}>M_{W'}(1-\epsilon/2)$ of the signal and background differential dijet mass distributions rather than their integrals in a finite interval centered on the $W'$ mass, because the former method is less sensitive to smearing effects generated by hadronization and jet reconstruction, which we cannot take into account in our parton-level analysis. In this way, we expect our estimate of the reach of the early LHC to be closer to the actual experimental results than it would be if we compared signal and background in an interval centered around the $W'$ mass.  
In addition to those in the dijet final state, also LHC searches in the $tb$ channel will be of course relevant to the $W'$ we are studying. We do not discuss them here, and refer the reader to the recent, extensive analysis of Ref.~\cite{Gopalakrishna:2010p2723}.

\subsection{Search for the $W'\rightarrow W\gamma$ decay} \label{LHCWgamma}

We now move on to consider decay channels of the $W'$ which have partial widths proportional to the $W$-$W'$ mixing angle $\hat{\theta}$. These include $WZ,\,Wh$ and $W\gamma$ final states. We will focus first on the last channel, which is of special interest since it is very suppressed in the gauge models containing a $(\mathbf{1},\mathbf{1})_{1}$ $W'$, such as for instance LR models. Therefore, observation of $W'\to W\gamma$ would point to a composite nature of the $W'$. The partial width for decay into $W\gamma$ reads
\begin{equation} \label{widthWgamma}
\Gamma(W'\rightarrow W\gamma)=\frac{e^{2}}{96\pi}(c_{B}+1)^{2}\sin^{2}\hat\theta\cos^{2}\hat\theta \left(1-\frac{M^{2}_{W}}{M^{2}_{W'}}\right)^{3}\left(1+\frac{M^{2}_{W'}}{M^{2}_{W}}\right)M_{W'}\,.
\end{equation}
Since the width for decay into this channel is controlled by $\hat\theta$ and $c_{B}$, it is interesting to estimate which values of these parameters will be accessible to the LHC in its first run. To assess the discovery potential, we choose two benchmark values for the $W'$ mass, namely 800 and 1200 GeV, and we assume two representative values of the integrated luminosity, namely 1 and 5 fb${}^{-1}$, at a center of mass energy of 7 TeV. We set the coupling to quarks to $g_{q}=0.84\,(1.48)g$ for $M_{W'}=800\,(1200)$ GeV, that is, to the largest value allowed by Tevatron $jj$ and $tb$ searches (see Fig.~\ref{fig:Tevatronlimits}). Notice that the upper limit on $g_{q}$ from Tevatron searches in quark final states was computed for $\hat{\theta}=0$; when the mixing is introduced, the bound on the coupling weakens, due to the smaller branching ratio of the resonance into quarks. 

A direct constraint on the mixing angle $\hat\theta$ comes from the non-observation of resonances decaying into $WZ$ in a search performed by the D0 collaboration \cite{D0Collaboration:2010p3231}: we take such constraint into account in our analysis for the $W\gamma$ final state. On the other hand, the CDF Collaboration has performed a search in the $\ell\gamma\MET$ ($\ell=e,\mu$) final state \cite{CDFCollaboration:2007p3228}, without observing any discrepancies with the SM prediction. Also the constraints coming from this channel were taken into account; however, they turn out to be less stringent than those obtained from the $WZ$ channel, because of the smaller dataset analyzed.

We select decays of the $W$ into an electron and a neutrino, and apply the following cuts on the $e\gamma\MET$ final state: $p_{T}^{\gamma}>250\,(400)$ GeV, $p_{T}^{e}>50$ GeV, $\MET >50\,\,\textrm{GeV}$, $|\eta_{e,\gamma}|<2.5$, and $|M(W\gamma)-M_{W'}|<0.05\,(0.10)M_{W'}\,$, for $M_{W'}=800\,(1200)$ GeV. We note that, even though the neutrino longitudinal momentum $p_{z}^{\,\nu}$ is not measured experimentally, it can be reconstructed by imposing that the lepton and neutrino come from an on-shell $W$: a quadratic equation for $p_{z}^{\,\nu}$ is thus obtained. It follows that a criterion must be chosen to unfold this ambiguity. The assessment of the effects of such choice on the cuts on $\MET$ and on the total invariant mass $M(W\gamma)$ goes beyond the scope of this work, and we leave it to the experimental collaborations\footnote{In this regard, we also note that, at the detector level, fluctuations in the measured $\MET$ can lead to events where no solution for $p_{z}^{\,\nu}$ can be found even though the lepton and neutrino come from the decay of a $W$ (see, e.g., the section on top quark mass measurements in Ref.~\cite{ATLASCollaboration:2008p3435}).}. We neglect the interference between $W$ and $W'$, which is due to the $O(\hat\theta)$ coupling of $W'$ to left-handed quark currents. The main background process is the SM $W\gamma$ production, which we include in our analysis, while we leave out the $W+j$ production with the jet misidentified as a photon. We have checked that applying the rejection factor for misidentification into a $\gamma$ of very high-$p_{T}$ jets, which is of the order of $5\times 10^{3}$ if photon identification and isolation cuts are applied (see, e.g., Ref.~\cite{ATLASCollaboration:2008p3435}), the $W+j$ background contribution is roughly one order of magnitude smaller than the irreducible $W\gamma$ process. This estimate suffers from the fact that we are not including NLO corrections to $W+j$, and from the fact that requiring photon identification and isolation has an efficiency of $\sim80\%$ on `real' photons \cite{ATLASCollaboration:2008p3435}, which would slightly reduce the number of signal events detected. Other possibly relevant instrumental backgrounds that we do not include in our exploratory study are $ee\MET$ with $e$ misidentified as a photon, and QCD jets faking $e+\MET$. We leave the proper treatment of such detector-dependent backgrounds to the experimental analyses; we just note that doubling the statistics by including also the $W\rightarrow \mu\nu$ channel would help in balancing the sensitivity loss, in case the sum of instrumental backgrounds -- such as those mentioned above -- happened to be of the same order of magnitude of the irreducible $W\gamma$ background (for example, in the D0 $\ell\gamma\MET$ search, the total background was estimated to be roughly twice as large as the irreducible $W\gamma$, see Ref.~\cite{CDFCollaboration:2007p3228}). 

\begin{figure}[t]
\centering
\begin{minipage}{20cm}
\includegraphics[scale=0.36]{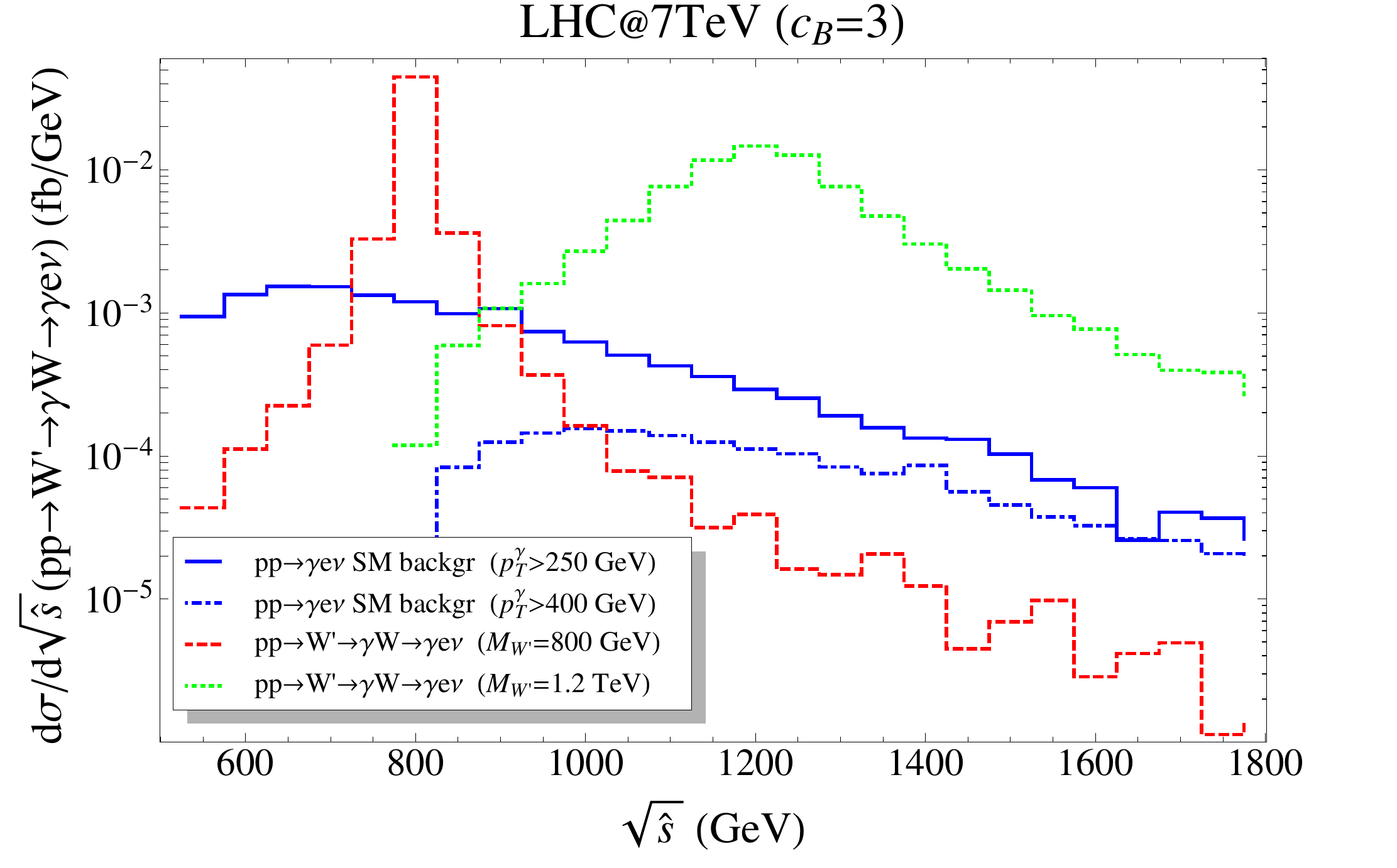}\hspace{-3mm}
\includegraphics[scale=0.36]{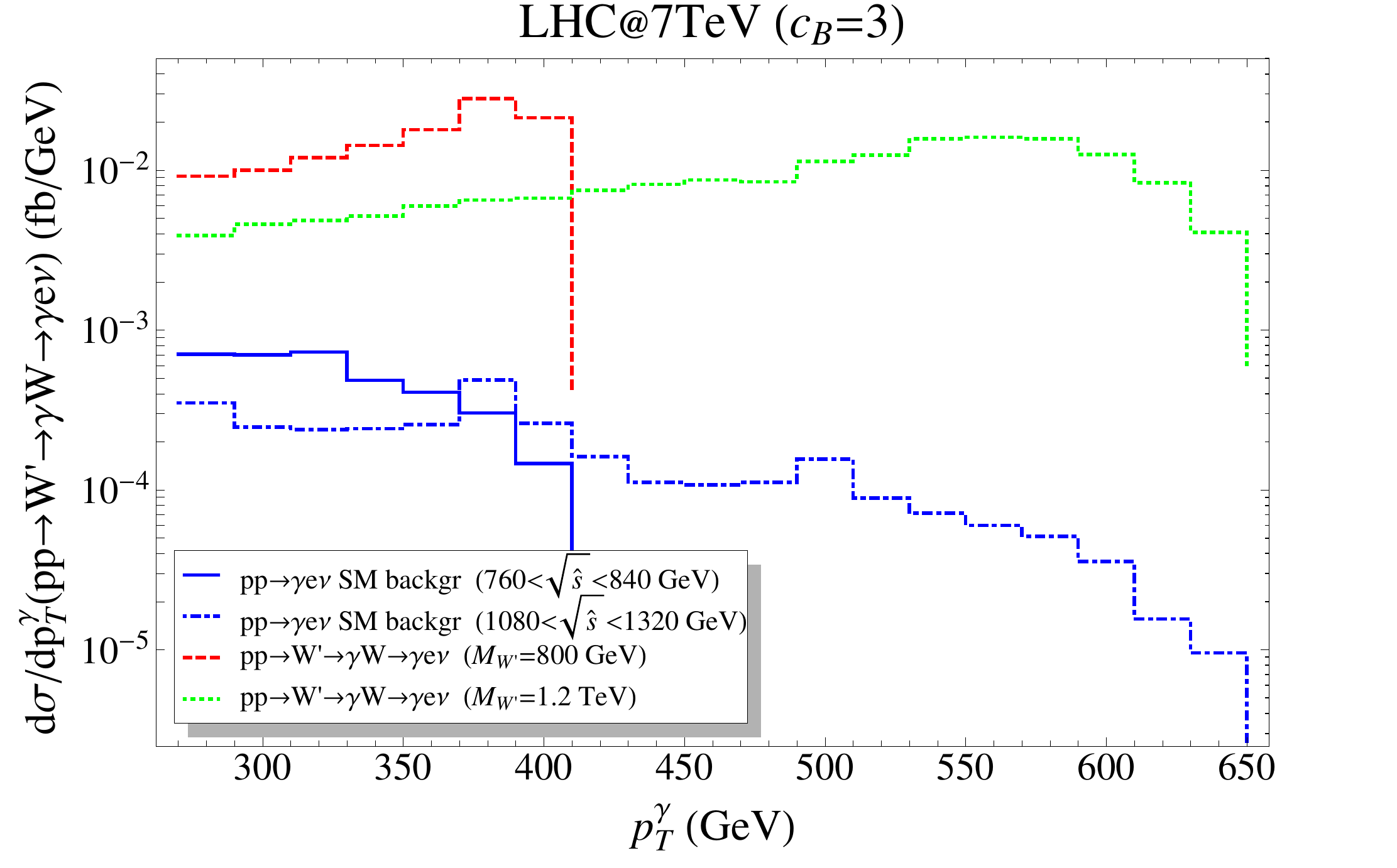}
\end{minipage}
\caption{Invariant mass (left) and photon $p_{T}$ (right) distributions for the \mbox{$W'\rightarrow W\gamma\rightarrow e\nu\gamma$} signal and for the irreducible background. The values of the couplings are as follows: \mbox{$g_{q}=0.84g$} and $\hat\theta = 10^{-2}$ for $M_{W'}=800\,\mathrm{GeV}$, and $g_{q}=1.48g$ and $\hat\theta =4\times 10^{-2}$ for $M_{W'}=1.2\,\mathrm{TeV}$.}
\label{fig:LHCWgammaDISTRIB}
\end{figure}

In Fig.~\ref{fig:LHCWgammaDISTRIB}, we show the distributions of the reconstructed invariant mass of the $W'$ and of the $p_{T}$ of the photon, compared to the SM $W\gamma$ background. We stress that experimentally, reconstruction of the longitudinal component of the neutrino momentum by imposing the on-shell condition for the $W$ will have an impact on the resolution of the $W'$ invariant mass. From the invariant mass distribution, it is also evident that for the values of the parameters chosen, the $W'$ of mass 1.2 TeV has a quite large width, which motivated the use of a broader cut around the peak, as discussed above. While the number of events predicted at the early LHC is clearly small, these distributions can be used as a guideline also for searches at higher integrated luminositites, after rescaling cross section to higher LHC center of mass energy. 

\begin{figure}[h]
\centering
\begin{minipage}{20cm}
\includegraphics[scale=0.37]{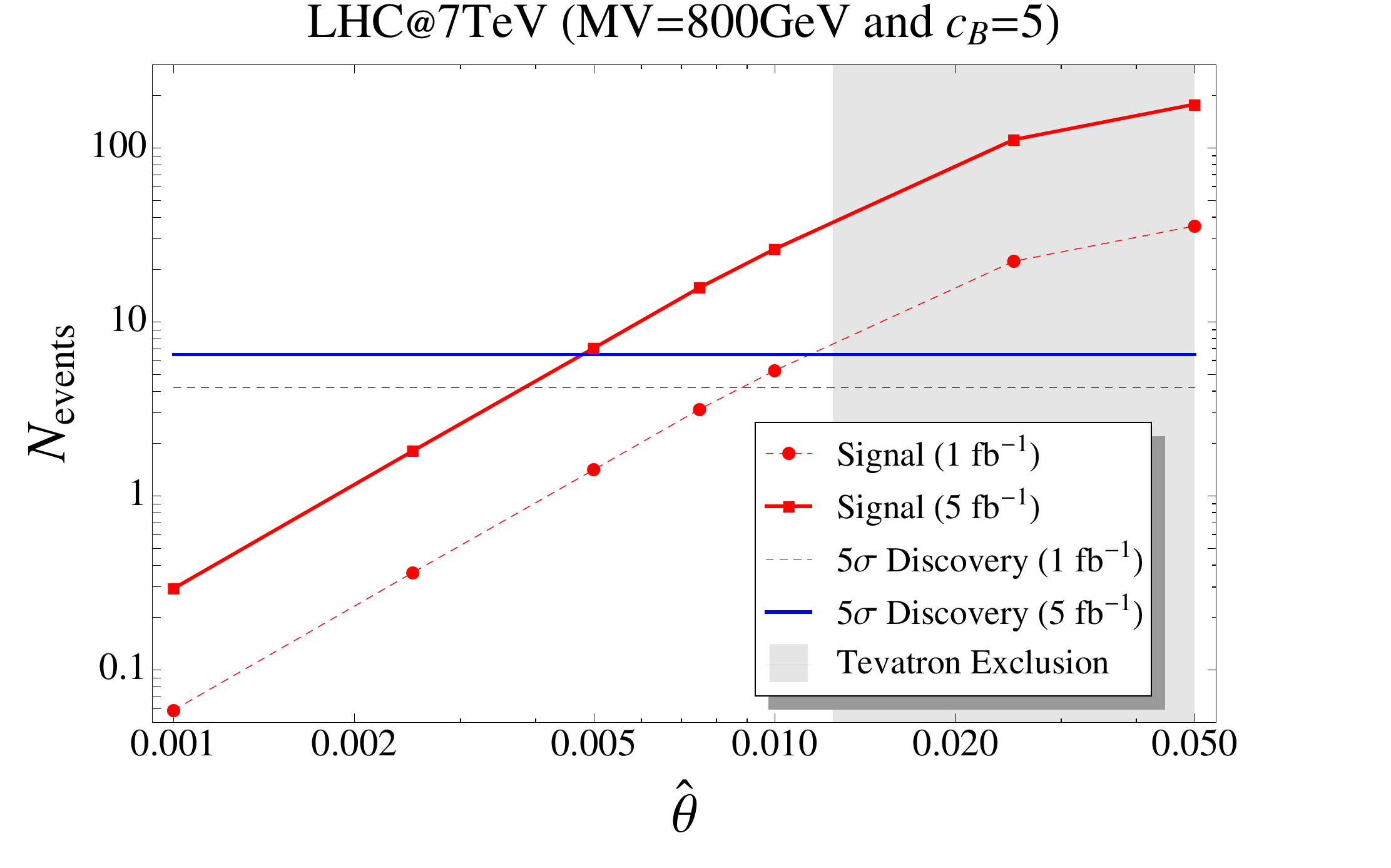}\hspace{-5mm}
\includegraphics[scale=0.37]{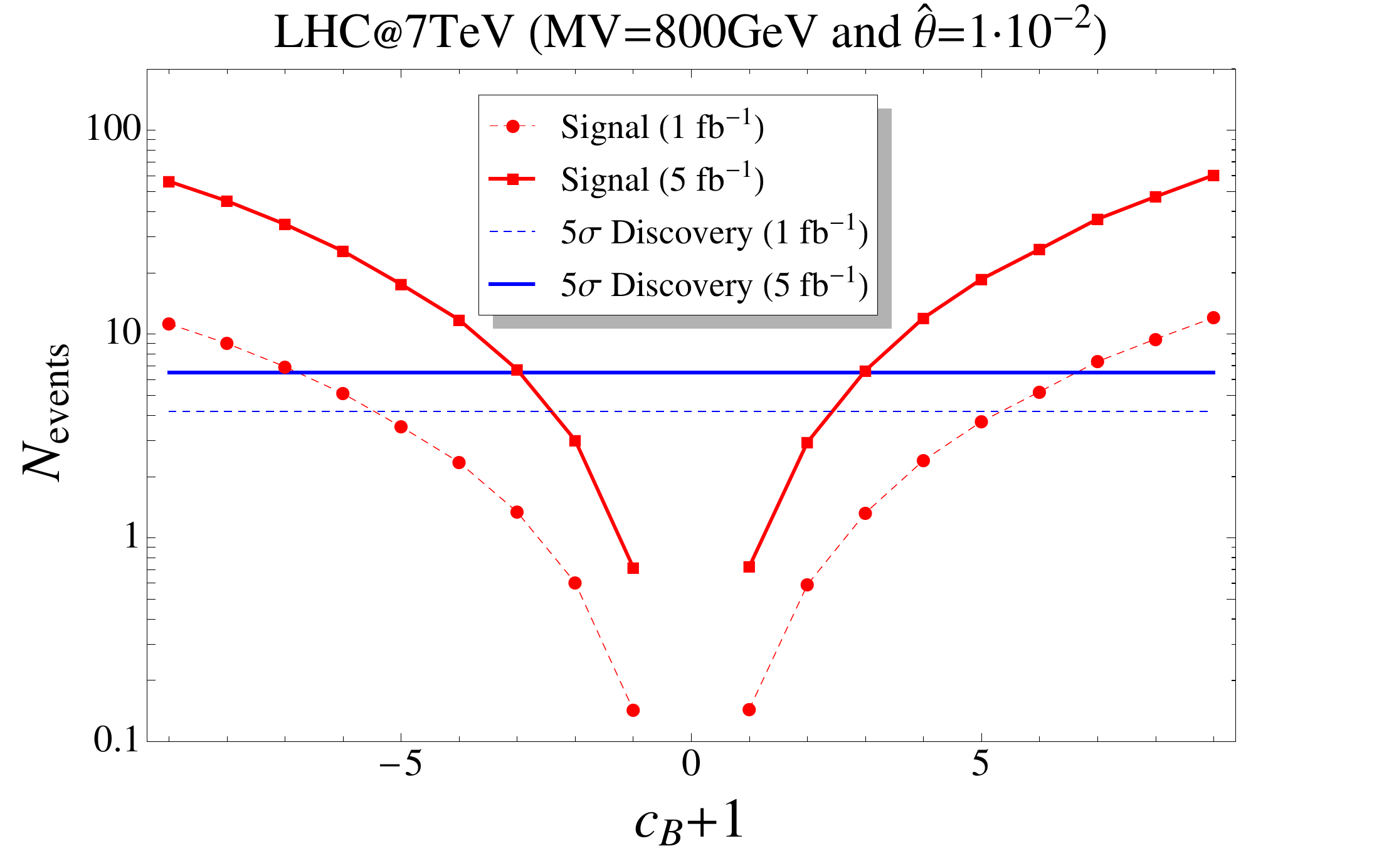} \\
\end{minipage}
\begin{minipage}{20cm}
\includegraphics[scale=0.37]{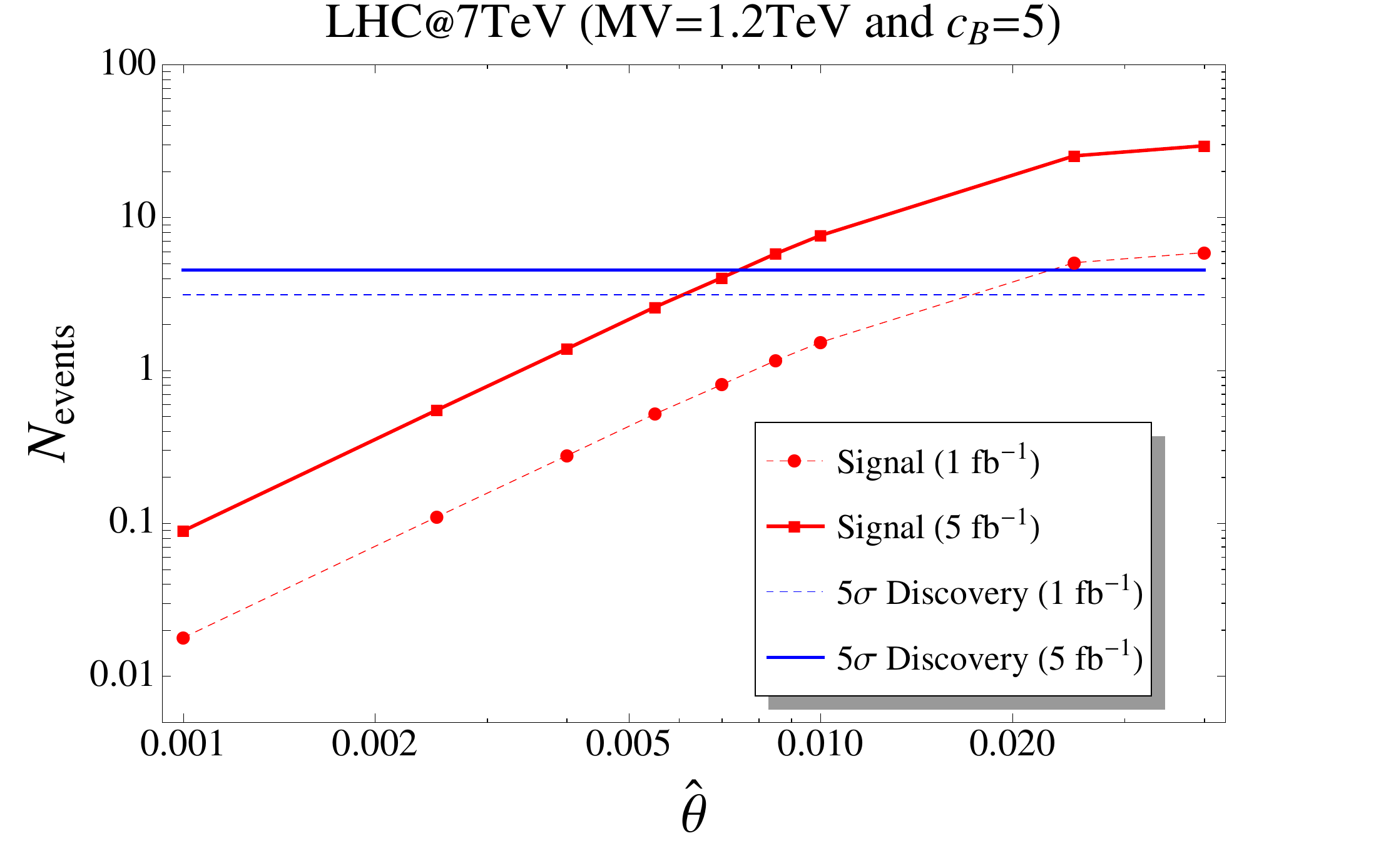} \hspace{-7mm}
\includegraphics[scale=0.37]{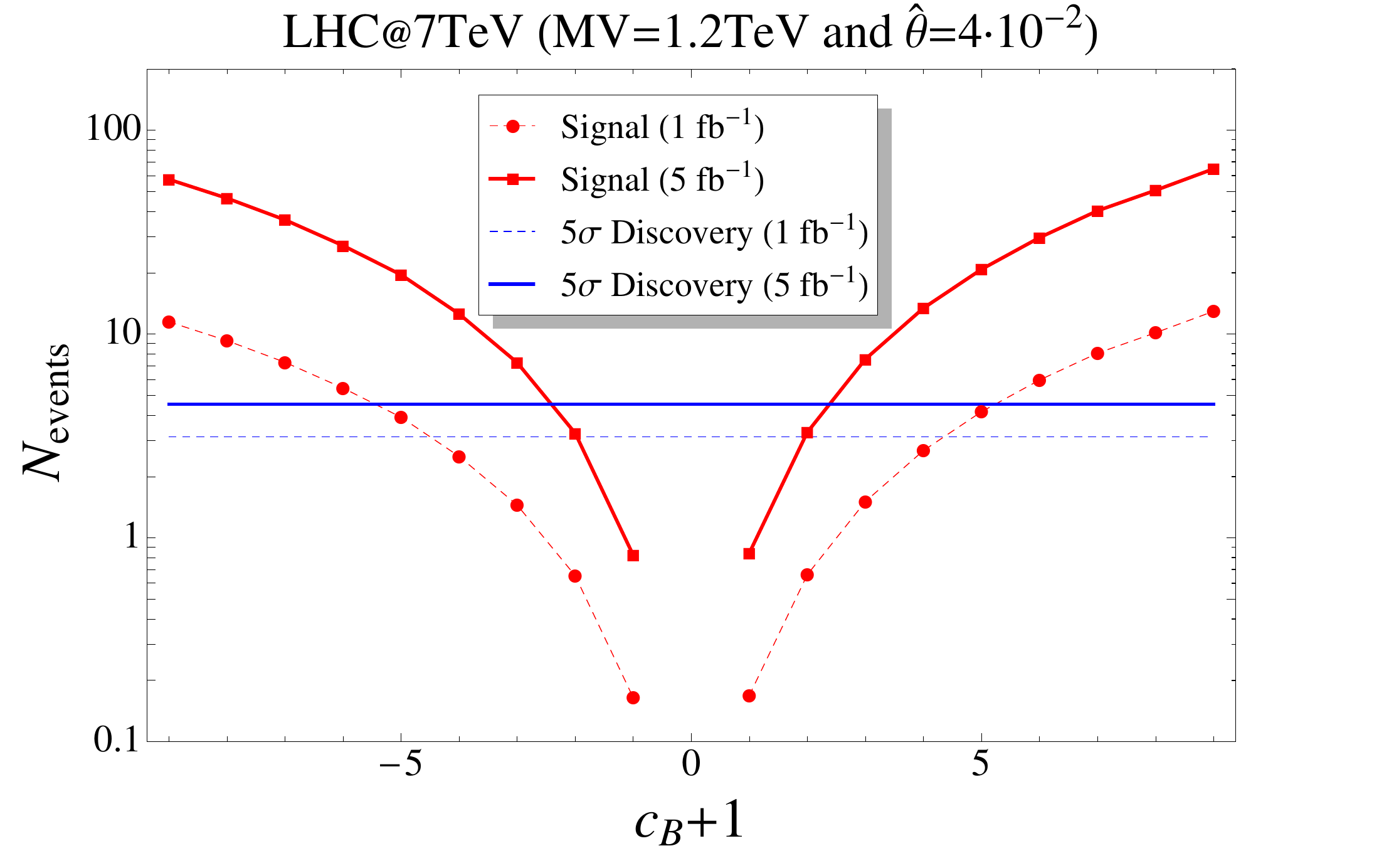}
\end{minipage}
\caption{`$5\sigma$' discovery prospects of the 7 TeV LHC for the $W'\rightarrow W\gamma\rightarrow e\nu \gamma$ process, for $M_{W'}=800$ GeV, $g_{q}=0.84g$ (top row) and $M_{W'}=1200$ GeV, $g_{q}=1.48g$ (bottom). $N_{\mathrm{events}}$ is the number of signal events after applying the cuts described in the text. The red curves show the expected number of events as a function of the parameters of our phenomenological Lagrangian, whereas the blue flat lines represent the number of events needed for a $5\sigma$ discovery, taking into account the SM background. The signal cross sections, after all cuts, are simply given by $\sigma_{S}=N_{\mathrm{events}}/L$; the background cross sections after all cuts are $\sigma_{B}\,(M_{W'}=800,1200\, \mathrm{GeV})= (9.6,\,2.7)\times 10^{-2}\,\mathrm{fb}$.  
The region shaded in grey is excluded at $95\%$ CL by Tevatron searches for resonances decaying into $WZ$.}
\label{fig:LHCWgamma}
\end{figure}

Our main results are shown in Fig.~\ref{fig:LHCWgamma}. As can be read off the left side of the figure, for $M_{W'}=800$ GeV, assuming $c_{B}=5$ (which corresponds to $g_{B}=5g'\sim 1.8$), the interval $5\times 10^{-3}<\hat\theta<1.25\times 10^{-2}$ is accessible for a discovery with 5 fb${}^{-1}$. Such values of $\hat\theta$, while being excluded by EWPT if we assume the $W'$ is the only new physics contributing to precision data (see the discussion after Eq.~\eqref{Tcontribution}), are however allowed by $u\rightarrow d$ and $u\rightarrow s$ transitions if the CP phases are not small. It is conceivable that a positive contribution to the $T$ parameter coming from additional new physics (such as, for example, a heavy neutral spin-1 state) relaxes the bound from EWPT, allowing for such relatively large values of $\hat\theta$. On the other hand, from the right side of Fig.~\ref{fig:LHCWgamma} we see that setting the mixing angle to the value $\hat\theta = 10^{-2}$, discovery of a $W'$ with mass \mbox{$800$ GeV} is possible with 5 fb$^{-1}$ for $c_{B}\gtrsim 2$, which corresponds to a moderate value of the coupling $g_{B}\sim 0.7$. The prospects for a heavier $M_{W'}=1200$ GeV are similar, except that in this case there is no relevant bound from Tevatron searches. 

For illustrative purposes, we also give in Table \ref{14TeV800GeV} an estimate of the sensitivity on $c_{B}$ for the 14 TeV LHC with 300 fb${}^{-1}$ luminosity. Background events are due to the irreducible SM $W\gamma$ process only. Cuts on the final state kinematics are the same as for the early LHC case discussed above.
\begin{table}[t]
\centering
\begin{tabular} {| c | c | c | c || c | c | c | c |}
\hline
$c_{B}+1$ & $N_{\mathrm{s}}$ & $N_{\mathrm{bckgr}}$ & $N_{\sigma}$ & $c_{B}+1$ & $N_{\mathrm{s}}$ & $N_{\mathrm{bckgr}}$ & $N_{\sigma}$  \\
\hline
$0.6$ & $57$  & $102$ & $5.7$ & $0.4$ & $34$  & $45$ & $5.0$ \\   
\hline
$0.5$ & $40$  & $102$ & $4.0$ & $0.3$ & $23$  & $45$ & $3.4$ \\   
\hline
$0.4$ & $26$  & $102$ & $2.6$ & $0.2$ & $9$  & $45$ & $1.5$ \\   
\hline
\end{tabular}
\caption{Sensitivity on $c_{B}$ at the 14 TeV LHC with 300 fb${}^{-1}$, for $M_{W'}=800$ GeV, $g_{q}=0.84g$ and $\hat\theta = 10^{-2}$ (left), and for $M_{W'}=1.2$ TeV, $g_{q}=1.48g$ and $\hat\theta = 4\times 10^{-2}$ (right).}
\label{14TeV800GeV}
\end{table}

Clearly, it is very interesting to understand what are the predictions for the strength of the $W'W\gamma$ coupling in extensions of the SM. In Ref.~\cite{Ferrara:1992yc} it was shown that the gyromagnetic ratio of any elementary particle of mass $M$ (of any spin) coupled to the photon has to take the value $g=2$, which can be equivalently written as $c_{B}=-1$ in our effective language, in order for perturbative unitarity to be preserved up to energies $E\gg M/e$. As a consequence, in any gauge extension of the SM, where the $W'$ is the fundamental gauge boson of some extra symmetry, $g=2$ has to be expected, since unitarity is preserved up to much larger scales. Indeed, in the `minimal' gauge model containing an isosinglet $W'$, namely a LR model (see App.~\ref{minimalgaugemodels} for the notation), we find that $c_{B}=-1$ at the renormalizable level. Including dimension-6 operators, we expect $c_{B}=-1+O(v_{R}^{2}/\Lambda^{2})$, where $\Lambda$ is the cut-off of the LR model. Therefore, $c_{B}\approx -1$ will still hold, and observation of $W'\to W\gamma$ is likely to be out of the reach of the LHC. 

On the other hand, if the $W'$ is a composite state of some new strong interaction, then the requirement of preservation of perturbative unitarity is relaxed, and significant departures from $c_{B}=-1$ can be envisaged. The only condition that needs to be satisfied even in the composite case is that the scale of violation of perturbative unitarity be sufficiently larger than the $W'$ mass. To verify that this is indeed the case, and since $c_{B}$ only appears in the $BVV$ vertex (see Eq.~\eqref{V-SM}), where $B$ is the hypercharge gauge boson and $V$ is the extra vector, we compute the amplitude for $BB\to VV$ scattering. The two independent amplitudes that grow the most with energy are $B_{+}B_{\pm}\to V_{L}V_{L}$, where $B_{\pm}$ are the two transverse polarizations of the $B$, and $V_{L}$ is the longitudinally polarized $V$. The leading term of these amplitudes in the high-energy limit reads
\begin{equation} \label{UVamplitudes}
A_{++\to LL}\approx \frac{(1-c_{B}^{2})g^{\prime\,2}s}{2M^{2}}\,,\qquad A_{+-\to LL}\approx \frac{(1+c_{B})^{2}g^{\prime\,2}s}{4M^{2}}\,. 
\end{equation} 
Notice that for $c_{B}\to -1$, the dangerous high-energy behavior is removed, as it was anticipated above. Requiring the amplitudes in Eq.~\eqref{UVamplitudes} not to exceed $16\pi^{2}$, we find the cut-off $\Lambda$ at which perturbative unitarity is lost\footnote{Other definitions of the perturbative unitarity bound are possible, and have been used in the literature. A different choice would simply change the numerical factors appearing in the definition of the cut-off.}, as a function of $c_{B}$: taking the maximum value we used in the phenomenological analysis, namely $c_{B}=10$, we find $\Lambda \approx 5 M$; for smaller values of $c_{B}$, the cut-off is obviously larger. This result guarantees that we can safely study the phenomenology at scale $M$ with relatively large values of $c_{B}$, without encountering any perturbative unitarity violation issues. 

We conclude that, since the size of the $W'W\gamma$ coupling is expected to be very small if the $W'$ is a fundamental gauge boson, observation of $W'\to W\gamma$ at the LHC would be a hint of the composite nature of the $W'$.    

\subsubsection{$W'\rightarrow W\gamma$ for a $W'$ belonging to an $SU(2)_{L}$ triplet}

It would be interesting to understand how the prospects in the search for \mbox{$W'\to W\gamma$} change if we consider a $W'$ transforming in the $(\mathbf{1},\mathbf{3})_{0}$ representation, which appears for example in some Little Higgs models and in models with large extra dimensions (the effective Lagrangian for such representation can be found in App.~\ref{EffLagr130}). Even though the $W'W\gamma$ interaction has the same structure for both the $(\mathbf{1},\mathbf{1})_{1}$ and the $(\mathbf{1},\mathbf{3})_{0}$ representations, see Eqs.~\eqref{W'Wgamma-111} and \eqref{W'Wgamma-130}, the results of our LHC study do not straightforwardly apply to the latter representation by just making the substitution \mbox{$c_{B}+1\,\to (c'_{B}+1)/(1-\widetilde{g}^{2})$}, because of $W$-$W'$ interference effects, which are potentially relevant for the isotriplet $W'$ (since it couples significantly to left-handed currents), and because of the different width (the triplet $W'$ also decays into light leptons). Furthermore, present constraints on the triplet $W'$ are different (and stronger) than those for the isosinglet $W'$ we consider in this paper. A detailed analysis of the $(\mathbf{1},\mathbf{3})_{0}$ $W'$ goes beyond the scope of this work.

From a theoretical perspective, in analogy with the isosinglet case, if the $SU(2)_{L}$ triplet $W'$ is a fundamental gauge boson (see App.~\ref{minimalgaugemodels} for the minimal gauge extension of the SM that contains such state), then $c'_{B}\approx -1$, and observation of the $W'\rightarrow W\gamma$ decay is likely to be out the LHC reach. On the other hand, if the $W'$ is a composite state, significant departures from $c'_{B}=-1$ are possible.

\subsection{Search for $W'\rightarrow WZ$} \label{WZsearch}

\begin{figure}[t]
\centering
\includegraphics[scale=0.37]{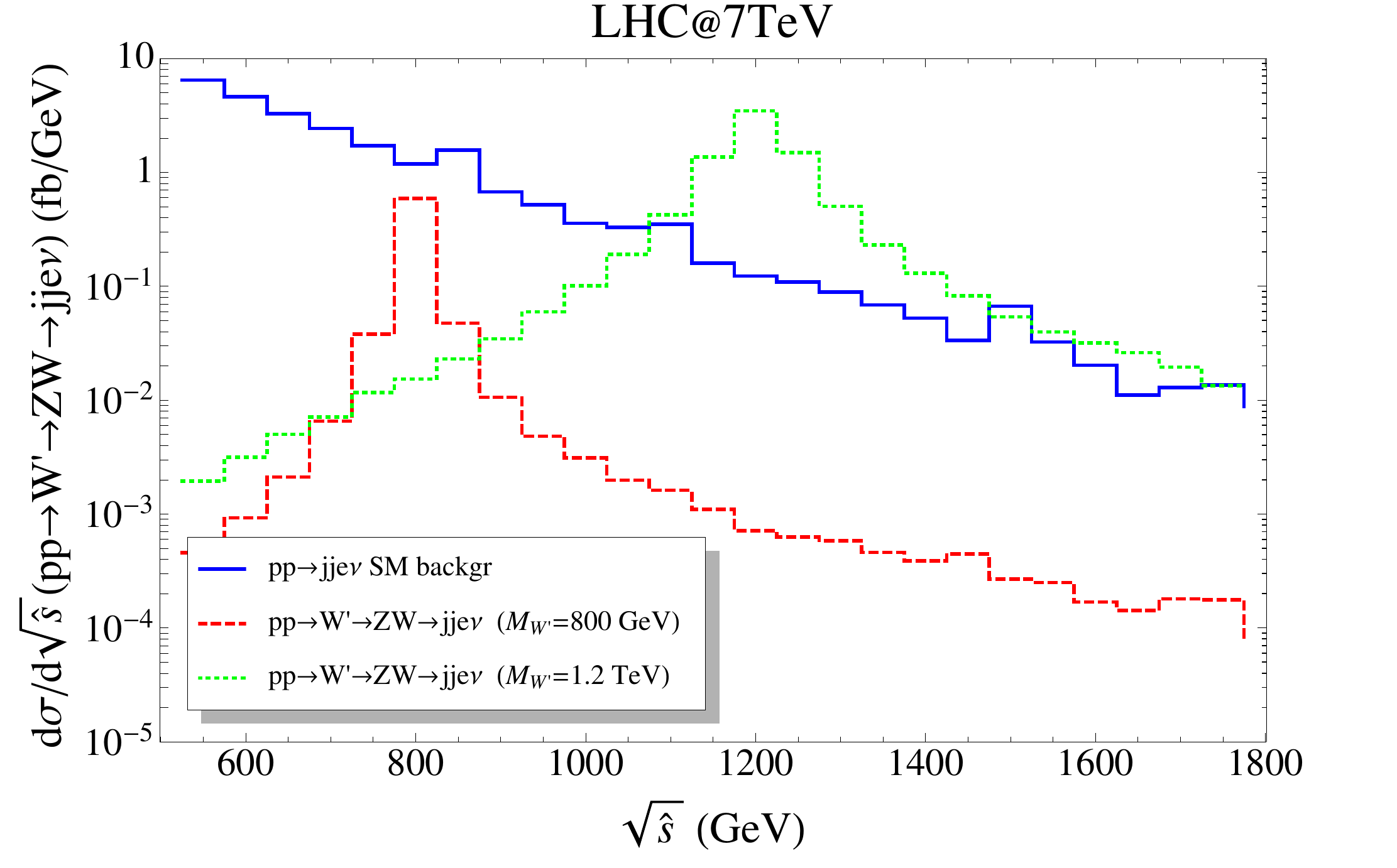}
\caption{Invariant mass distribution of the $e\nu jj$ system for the $W'$ signal and for the $pp\rightarrow e\nu jj$ background. The values of the coupling $g_{q}$ are the same as in Fig.~\ref{fig:LHCWgammaDISTRIB}.}
\label{fig:WZdistrib}
\end{figure}

\begin{figure}[t]
\centering
\begin{minipage}{20cm}
\includegraphics[scale=0.37]{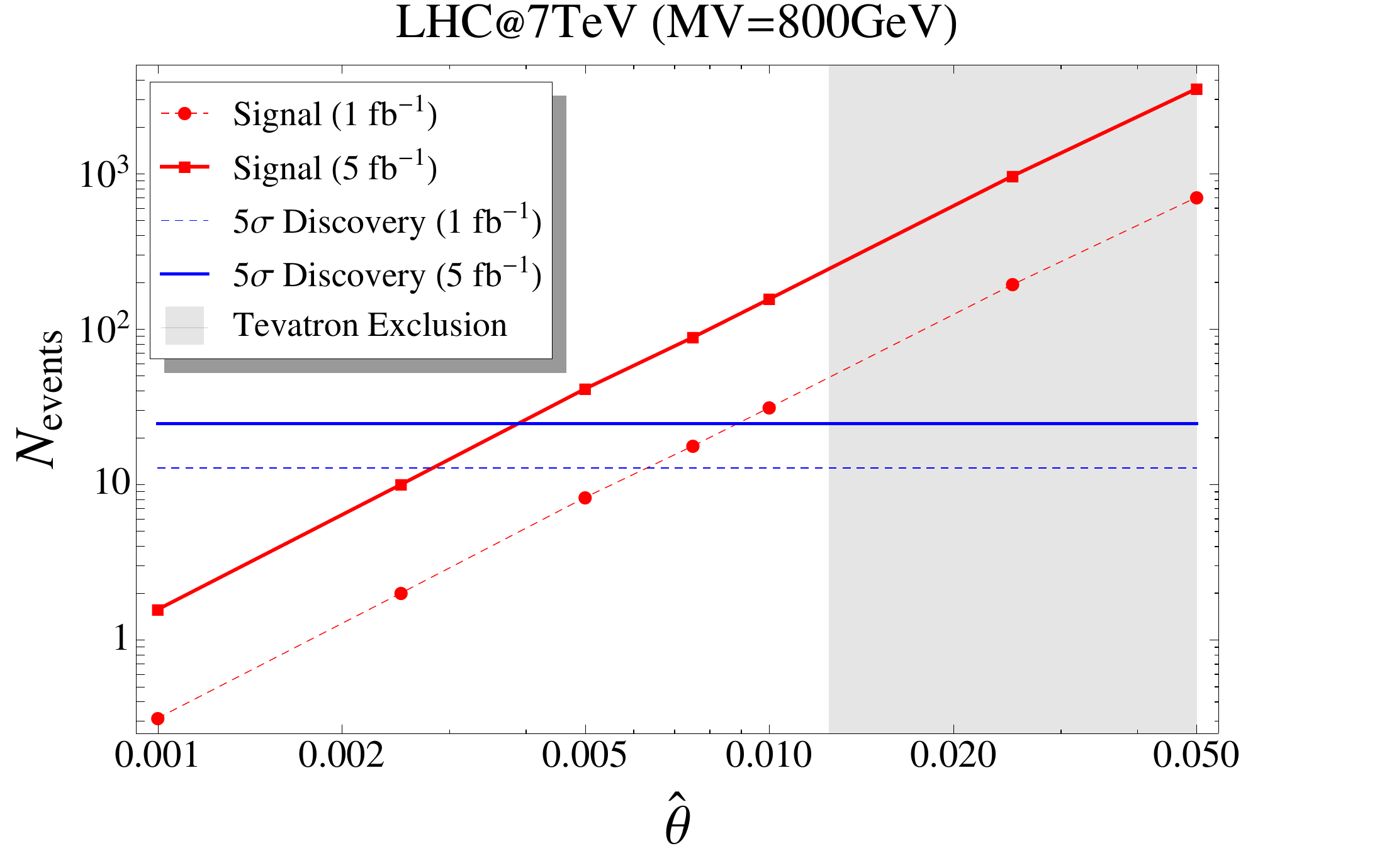}\hspace{-5mm}
\includegraphics[scale=0.37]{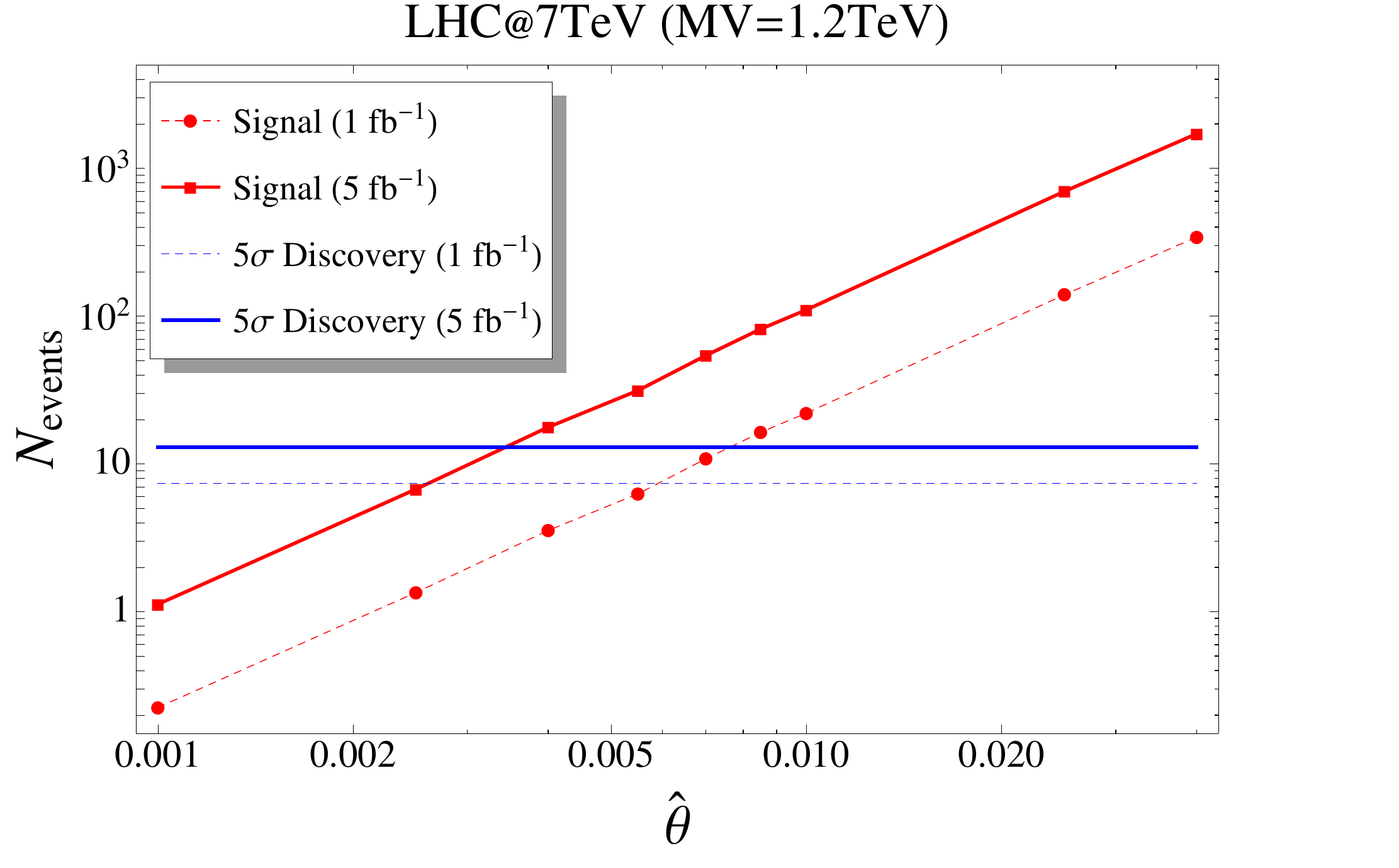}
\end{minipage}
\caption{`$5\sigma$' discovery prospects on the mixing angle $\hat\theta$ via the $W'\rightarrow WZ\rightarrow e\nu jj$ process at the 7 TeV LHC, for $M_{W'}=800\,\mathrm{GeV}$, $g_{q}=0.84g$ (left) and $M_{W'}=1200\,\mathrm{GeV}$, $g_{q}=1.48g$ (right). The results are almost independent of $c_{B}$. The interpretation of the curves is analogous to Fig.~\ref{fig:LHCWgamma}; after all cuts, the background cross sections are $\sigma_{B}\,(M_{W'}=800,1200\, \mathrm{GeV})= (3.5,\,0.73)\,\mathrm{fb}$. The region shaded in grey is excluded at $95\%$ CL by Tevatron searches for resonances decaying into $WZ$.}
\label{fig:WZ}
\end{figure}

We also discuss the $W'\rightarrow WZ$ decay, which is complementary to $W'\rightarrow W\gamma$ because, being the rate for resonant $WZ$ production almost independent of the parameter $c_{B}$, its measurement would allow one to estimate the size of the mixing angle $\hat\theta$. Since we consider the early LHC reach, where integrated luminosity will be $\lesssim 5$ fb$^{-1}$, the most promising final state is $WZ\rightarrow \ell \MET jj$, which has a larger rate with respect to the purely leptonic channel; on the contrary, selecting leptonic decays of the $Z$ together with a hadronic $W$ has been shown to be less promising \cite{Alves:2009aa}. 
Therefore, we implement simple cuts on the $e\nu jj$ final state (we only consider $W$ decays into an electron, in analogy to what we did for the $W'\rightarrow W\gamma$ process) to enhance the ratio of signal over background, namely: $p_{T}^{e,j}>50$ GeV, $\MET > 50$ GeV, $|\eta_{e,j}|<2.5$, and in addition we require the invariant mass of the dijet system to reconstruct a $Z$, $|M(jj)-M_{Z}|<20\,\mathrm{GeV}$. Finally, we select events which have an invariant mass compatible with $M_{W'}$ as follows: $|M(e\nu jj)-M_{W'}|<0.10M_{W'}\,$. The background we consider is the SM $pp\rightarrow e\nu jj$, which includes a large contribution from $W+jj$. The $t\overline{t}$ background can be efficiently reduced to roughly one order of magnitude less than the QCD background by applying a central jet veto \cite{Alves:2009aa}, and we do not consider it here. The invariant mass distributions of signal and background for this channel are shown in Fig.~\ref{fig:WZdistrib}.
Our results are shown in Fig.~\ref{fig:WZ} for the same choices of the $W'$ mass and couplings that we already discussed when studying $W'\rightarrow W\gamma$, so that a direct comparison between the two searches can be made. We can see that with 5 fb${}^{-1}$, a mixing angle larger than $\hat\theta\approx 3\,\text{--}\, 4\times 10^{-3}$ is accessible for discovery; this result is to a good approximation independent of the size of $c_{B}$. We also notice that the number of signal events can be sizable, which is the main reason why this channel is more favorable than the purely leptonic one for limited LHC luminosity. We remark that the size of the cut on the total invariant mass of $e\nu jj$ agrees with Ref.~\cite{Alves:2009aa}, where it was chosen to retain most of the signal in the presence of jet energy smearing. In addition, the cut we set on the invariant mass of the $jj$ system is even looser than the one adopted in Ref.~\cite{Alves:2009aa}. Therefore we believe our results to be reasonably stable with respect to jet smearing, which was not included in our parton-level analysis. 

Finally, we do not discuss $W'$ decays into $Wh$, because the choice of the most relevant final states is strongly dependent on the Higgs mass, and such a detailed study goes beyond the scope of this work. We refer the interested reader to Refs.~\cite{Azuelos:2004dm,Bao:2011nh} and to the references cited therein. 


\section{Summary and conclusions} \label{summary}

We have applied an effective approach to study the phenomenology of a heavy $W'$ transforming as a singlet under weak isospin. Such a $W'$ is very weakly coupled to light leptons, and is therefore only constrained by $\Delta F=2$ hadronic processes (mainly $K$ and $B$ meson mixing) and Tevatron direct searches, provided the $W$-$W'$ mixing angle $\hat\theta$ is small enough to evade the important bounds from the oblique $T$ parameter and from precision measurements of $u\rightarrow d$ and $u\rightarrow s$ semileptonic transitions.\footnote{We remark that in the effective approach, the coupling to quarks and the mixing angle are independent parameters.} Furthermore, for suitable choices of the right-handed quark mixing matrix $V_{R}$, the only constraints on the coupling of the $W'$ to quarks come from Tevatron direct searches. Therefore, a $W'$ with mass even below a TeV and sizable coupling to quarks is allowed by present data. We have estimated the early LHC reach in the dijet channel on such a resonance. We have also noted that, if different choices for $V_{R}$ are made, our effective approach encompasses the class of $W'$ with flavor-violating couplings that has been recently called for as an explanation of the anomaly observed by CDF in the top pair forward-backward asymmetry.

Subsequently we have discussed the possibility that the $W$-$W'$ mixing angle be large enough to allow observation of the decays $W'\rightarrow W\gamma$ and $W'\to WZ$ at the early LHC. We have shown that discovery of these decays is possible for values of $\hat\theta$ allowed by semileptonic processes, if the CP phases in $V_{R}$ are not small. Although such values of $\hat\theta$ are excluded by EWPT because of the too large negative contribution the $W'$ gives to the $T$ parameter, it is conceivable that some additional new physics, such as for example an additional heavy neutral vector, could relax such constraint. 

We have shown that the $W'\to W\gamma$ channel is of significant relevance to gain insight on the nature of the $W'$ after a discovery in the dijet (or $tb$) final state. We have compared the experimentally accessible values of the parameters $(c_{B},\hat\theta)$ to the prediction for the strength of the $W'W\gamma$ coupling both in weakly coupled gauge extensions of the SM, and in strongly interacting theories where the $W'$ is a composite state. We have shown that observation of $W'\rightarrow W\gamma$ at the early LHC would be a hint of the composite nature of the resonance. We also briefly commented on the relevance of the decay into $W\gamma$ in case the $W'$ belongs to a triplet under weak isospin, the other representation which is commonly encountered in BSM constructions.  


\acknowledgments
We thank G.~Isidori, D.~Pappadopulo, G.~Villadoro and F.~Zwirner for discussions. We are grateful to M.~Narain for clarifications about Ref.~\cite{Abazov:2011xs}. This research has been partly supported by the European Commission under the contract ERC Advanced Grant 226371 {\it MassTeV} and the contract PITN-GA-2009-237920 {\it UNILHC}. The work of R.~T. has been supported in part by the European Commission under the contract PITN-GA-2010-264564 {\it LHCPhenoNet}. E.~S. was partly supported by the Fondazione Cariparo Excellence Grant \textit{String-derived supergravities with branes and fluxes and their phenomenological implications}.

\appendix

\section{Partial decay widths} \label{Partialdecaywidths}
Defining
\begin{equation*}
p=\frac{1}{2M_{W'}}\sqrt{M_{W'}^{4}+M_{1}^{4}+M_{2}^{4}-2M_{W'}^{2}M_{1}^{2}-2M_{W'}^{2}M_{2}^{2}-2M_{1}^{2}M_{2}^{2}}\,,
\end{equation*}
with $M_{1,2}$ the masses of the final state particles, the two body widths are given below.\\
The decay width into a pair of quarks reads
\begin{align} \nonumber
\Gamma(W^{\prime\,+}\rightarrow u^{i}\overline{d}^{j})=& \frac{p}{2\pi M_{W'}^{2}}\Big[|(v^{\prime})_{ij}|^{2}(3\sqrt{m_{d}^{2}+p^{2}}\sqrt{m_{u}^{2}+p^{2}}+3m_{d}m_{u}+p^{2}) \\
& +|(a^{\prime})_{ij}|^{2}(3\sqrt{m_{d}^{2}+p^{2}}\sqrt{m_{u}^{2}+p^{2}}-3m_{d}m_{u}+p^{2})\Big]\,,
\end{align}
while the decay width into two leptons, neglecting their masses, is
\begin{align} \nonumber
\Gamma(W^{\prime\,+}\to \nu^{i}\overline{e}^{i})=\frac{M_{W'}}{48\pi}g^{2}\sin^{2}\hat{\theta}\,.
\end{align}
The width for decay into a $W$ and a photon is reported in Eq.~\eqref{widthWgamma}, whereas for decay into a $W$ and a Higgs we find
\begin{equation}
\Gamma(W^{\prime}\rightarrow Wh)=\frac{p}{8\pi M_{W'}^{2}}\frac{v^{2}}{3}K^{2}\left(3+\frac{p^{2}}{M_{W}^{2}}\right)\,,
\end{equation}
where 
$$
K=\frac{g_{4}^{2}-g^{2}}{2}\sin\hat{\theta}\cos\hat\theta+\frac{g_{H}g}{\sqrt{2}}(\cos^{2}\hat\theta-\sin^{2}\hat\theta)\,.
$$
Finally,
\begin{equation}
\Gamma(W^{\prime}\rightarrow WZ)=\frac{p}{8\pi M_{W'}^{2}}\frac{g^2\cos^{2}\theta_{w}}{3}\sin^{2}\hat\theta\cos^{2}\hat\theta\times \mathcal{T}(M_{W'}^{2},M^{2}_{Z},M^{2}_{W}; c_{B})\,,
\end{equation}
where
\small\begin{align*}
\mathcal{T}&(M_{W'}^{2},M^{2}_{Z},M^{2}_{W}; c_{B})= \frac{1}{M_W^2 M_Z^2}p^2 \Bigg\{\tan^2\theta_{w} \Bigg[\tan^2\theta_{w} \Big[c_B^2 M_Z^2 (4 E_W E_Z+3 M_Z^2+4 p^2)\\ &+6 M_{W'} (E_W+E_Z)(-c_B M_Z^2+M_W^2)+M_W^2 (2 (-c_B (-2 c_B+3)+2) M_Z^2+4 E_W E_Z+4 p^2)\\ &+M_{W'}^2 (2 E_W E_Z+3 M_W^2+3 M_Z^2+2 p^2)+3 M_W^4\Big]+2 \Big[3 M_{W'} (E_W+E_Z) ((-c_B+1) M_Z^2\\ &+2 M_W^2)+M_W^2 (7 (-c_B+1) M_Z^2 +4 E_W E_Z+4 p^2)-c_B M_Z^2 (4 E_W E_Z+3 M_Z^2+4 p^2)\\ &+M_{W'}^2 (2 E_W E_Z+3 M_W^2+3 M_Z^2+2 p^2) +3 M_W^4\Big]\Bigg]+M_{W'}^2 \Big[2 (E_W E_Z+p^2)+3 M_W^2\\ &+3 M_Z^2\Big]+6 M_{W'} (E_W+E_Z) (M_W^2+M_Z^2) +2 M_W^2 \Big[2 (E_W E_Z+p^2)+7 M_Z^2\Big]\\ &+M_Z^2 \Big[4 (E_W E_Z+p^2)+3 M_Z^2\Big]+3 M_W^4\Bigg\}\,,
\end{align*}\normalsize
and $E_{W,Z}\equiv\sqrt{M^{2}_{W,Z}+p^{2}}$. In the limit $M_{W'}\gg M_{W,\,Z}$, this simplifies to
\begin{equation*}
\mathcal{T}(M^{2}_{W'}\gg M_{W,\,Z}^{2})\approx \frac{M^{6}_{W'}}{4M^{2}_{Z}M^{2}_{W}}(1+\tan^{2}\theta_{w})^{2}\,,
\end{equation*}
which is independent of $c_{B}$.

\section{Effective Lagrangian for the $(\mathbf{1},\mathbf{3})_{0}$ representation} \label{EffLagr130}
The effective Lagrangian for an $SU(2)_{L}$ triplet reads:
\begin{equation}
\mathcal{L}=\mathcal{L}_{SM}+\mathcal{L}_{V}+\mathcal{L}_{V-SM}\,,
\end{equation}
where $\mathcal{L}_{SM}$ is the SM Lagrangian, while 
\begin{equation}
\mathcal{L}_{V}=-\frac{1}{4}V^{\mu\nu\,a}V_{\mu\nu}^{a}+\frac{1}{2}M^{2}V_{\mu}^{a}V^{\mu\,a}
\end{equation}
with $V^{\mu\nu\,a}= \partial^{\mu}V^{\nu\,a}-\partial^{\nu}V^{\mu\,a}+ g \epsilon^{abc}(\hat{W}_{\mu}^{b}V_{\nu}^{c}-\hat{W}_{\nu}^{b}V_{\mu}^{c})$.
On the other hand,
\begin{equation} \nonumber
\mathcal{L}_{V-SM}=c'_{B}\frac{g}{2}\epsilon^{abc}\hat{W}_{\mu\nu}^{a}V^{\mu\,b}V^{\nu\,c}+g_{V}\epsilon^{abc}V_{\mu\nu}^{a}V^{\mu\,b}V^{\nu\,c}+\frac{g_{4}^{\prime\,2}}{4} V_{\mu}^{a}V^{\mu\,a}|H|^{2}+\frac{\widetilde{g}}{2}V^{\mu\nu\,a}\hat{W}_{\mu\nu}^{a}
\end{equation}
\begin{equation}
+V_{\mu}^{a}(g_{V}^{l})_{ij}\overline{l_{L}^{i}}\gamma^{\mu}\frac{\sigma^{a}}{2}l_{L}^{j}+V_{\mu}^{a}(g_{V}^{q})_{ij}\overline{q_{L}^{i}}\gamma^{\mu}\frac{\sigma^{a}}{2}q_{L}^{j}+(iV_{\mu}^{a}g'_{H}H^{\dagger}\frac{\sigma^{a}}{2}D^{\mu}H+\text{h.c.})\,,
\end{equation}
where
$
\hat{W}^{\mu\nu\,a}=\partial^{\mu}\hat{W}^{\nu\,a}-\partial^{\nu}\hat{W}^{\mu\,a}+g\epsilon^{abc}\hat{W}^{\mu\,b}\hat{W}^{\nu\,c}\,.
$
We denote with a hat the $SU(2)_{L}$ gauge bosons in this basis. Similarly to the $(\mathbf{1},\mathbf{1})_{1}$ case, we neglect operators that would only contribute to quartic interactions of spin-1 fields. The kinetic terms are made canonic by means of the following transformation:
\begin{equation} \label{kinetic transf}
\begin{pmatrix}
\hat{W} \\
V \end{pmatrix} =\begin{pmatrix}
1 & \frac{\widetilde{g}}{\sqrt{1-\widetilde{g}^{2}}} \\
0 & \frac{1}{\sqrt{1-\widetilde{g}^{2}}} \end{pmatrix}
\begin{pmatrix}
\bar{W} \\
\bar{V} \end{pmatrix}\,.
\end{equation}
After performing the transformation \eqref{kinetic transf} on the Lagrangian, the neutral and charged mass matrices are diagonalized by the following transformations:
\begin{align}
\begin{pmatrix} \bar{W}^{3} \\
B \\
\bar{V}^{3} \end{pmatrix} =&\, \begin{pmatrix}
\sin\theta_{w} & \cos\theta_{w} & 0 \\
\cos\theta_{w} & -\sin\theta_{w} & 0 \\
0 & 0 & 1 \end{pmatrix}
\begin{pmatrix}
1 & 0 & 0 \\
0 & \cos\theta_{n} & -\sin\theta_{n} \\
0 & \sin\theta_{n} & \cos\theta_{n} \end{pmatrix}
\begin{pmatrix}
A \\
Z \\
Z' \end{pmatrix}\,, \\
\begin{pmatrix}
\bar{W}^{+} \\
\bar{V}^{+} \end{pmatrix} =&\, \begin{pmatrix}
\cos\theta_{c} & -\sin\theta_{c} \\
\sin\theta_{c} & \cos\theta_{c} \end{pmatrix}
\begin{pmatrix}
W^{+} \\
W^{\prime\,+} \end{pmatrix}\,.
\end{align} 
Here $\theta_{w}$ is as usual the weak mixing angle, whereas $\theta_{n},\theta_{c}$ are the $Z$-$Z'$ and $W$-$W'$ mixing angles respectively. Their expressions read 
\begin{equation*}
\tan(2\theta_{n})=\frac{2\Delta^{2}_{\bar{Z}\bar{V}^{3}}}{M^{2}_{\bar{Z}}-M^{2}_{\bar{V}^{3}}}\,,\qquad \tan(2\theta_{c})=\frac{2\Delta^{2}_{\bar{W}^{+}\bar{V}^{-}}}{M^{2}_{\bar{W}^{+}}-M^{2}_{\bar{V}^{+}}}\,,
\end{equation*}
where
\begin{equation*}
M^{2}_{\bar{V}^{3}}=M^{2}_{\bar{V}^{+}}=\frac{1}{1-\widetilde{g}^{2}}\left(M^{2}+\frac{v^{2}g_{4}^{\prime\,2}}{4}+\frac{v^{2}g\widetilde{g}(2g_{H}+g\widetilde{g})}{4}\right)\,,
\end{equation*}
\begin{equation*}
\Delta^{2}_{\bar{Z}\bar{V}^{3}}=\frac{\sqrt{g^{2}+g^{\prime\,2}}(g_{H}+g\widetilde{g})v^{2}}{4\sqrt{1-\widetilde{g}^{2}}}\,,\qquad 
\Delta^{2}_{\bar{W}^{+}\bar{V}^{-}}=\frac{g(g_{H}+g\widetilde{g})v^{2}}{4\sqrt{1-\widetilde{g}^{2}}}\,,
\end{equation*}
and $M^{2}_{\bar{Z}}=(1/4)(g^{2}+g^{\prime\,2})v^{2}\,,\,\,M^{2}_{\bar{W}^{+}}=(1/4)g^{2}v^{2}\,$.

Once we write the Lagrangian in the mass eigenstate basis, the $W'W\gamma$ interaction reads
\begin{equation} \label{W'Wgamma-130}
-ie\sin\theta_{c}\cos\theta_{c}F^{\mu\nu}(W^{+}_{\mu}W^{\prime\,-}_{\nu}+W^{-}_{\nu}W_{\mu}^{\prime\,+})\left(\frac{1+c'_{B}}{1-\widetilde{g}^{2}}\right)\,.
\end{equation}
If the $W'$ is a gauge boson, then $c'_{B}=-1$ at the renormalizable level. This is completely analogous to what we discussed for a $W'$ in the $(\mathbf{1},\mathbf{1})_{1}$ representation.

\section{Minimal gauge models containing a $(\mathbf{1},\mathbf{1})_{1}$ or $(\mathbf{1},\mathbf{3})_{0}$ $W'$} \label{minimalgaugemodels}

In this appendix we set the notation for the `minimal' gauge models which contain in their spectrum an isosinglet or isotriplet $W'$, namely LR models and models based on $SU(2)_{1}\times SU(2)_{2}\times U(1)_{Y}$, respectively. It is easy to verify that in both cases, the $W'W\gamma$ vertex is vanishing at the renormalizable level, as it is expected in general in gauge models (see the discussion in Section~\ref{LHCWgamma}).

\subsection{$SU(2)_{L}\times SU(2)_{R}\times U(1)_{B-L}$ model} 
We consider an `asymmetric' LR model, based on the group $SU(2)_{L}\times SU(2)_{R}\times U(1)_{X}$, $X=(B-L)/2$, which is the simplest gauge extension of the SM containing a vector in the $(\mathbf{1},\mathbf{1})_{1}$ representation. By asymmetric, we mean that we do not assume any discrete symmetry relating the left and right sectors: in particular, $g_{R}\neq g_{L}$ in general. The breaking $SU(2)_{R}\times U(1)_{B-L}\rightarrow U(1)_{Y}$ is realized by a doublet\footnote{We employ the notation $(SU(2)_{L},SU(2)_{R},(B-L)/2)$ to label the representation.} $H_{R}\sim (1,2,1/2)$, with \textsc{vev}
$$
\left\langle H_{R}\right\rangle=\frac{1}{\sqrt{2}}\begin{pmatrix}
0 \\
v_{R}
\end{pmatrix}\,.
$$
The hypercharge coupling is identified as $1/g^{\prime\,2}=1/g_{R}^{2}+1/g_{X}^{2}$. Electroweak symmetry is broken by a bi-doublet $\Phi\sim (2,2,0)$, and we also consider a doublet $H_{L}\sim (2,1,1/2)$. With a generic Higgs potential, the \textsc{vev}s of these fields can be written as
$$
\left\langle\Phi\right\rangle = \begin{pmatrix}
k\,\, & 0 \\
0\,\, & k'e^{i\alpha_{1}} 
\end{pmatrix}\,, \qquad \left\langle H_{L}\right\rangle=\frac{1}{\sqrt{2}}\begin{pmatrix}
0 \\
v_{L}e^{i\alpha_{2}}
\end{pmatrix}\,.
$$
In the charged sector, the \textsc{vev} of $\Phi$ generates a mass mixing between $W_{L}$ and $W_{R}$,
$$
\begin{pmatrix}
W_{L}^{+}\\
W_{R}^{+}
\end{pmatrix}= \begin{pmatrix}
\cos\xi\,\, & -\sin\xi \\
e^{i\alpha_{1}}\sin\xi\,\, & e^{i\alpha_{1}}\cos\xi \end{pmatrix}
\begin{pmatrix}
W^{+} \\
W^{\prime\,+}
\end{pmatrix},
$$
with $\xi\sim kk'/v_{R}^{2}$. In the neutral sector, diagonalization of the mass matrix is obtained through the rotation (we take for simplicity the limit $v_{L}\to 0$)
\begin{equation}
\begin{pmatrix}
W^{3}_{L} \\
W^{3}_{R}\\
X \end{pmatrix}
= \begin{pmatrix}
1 & 0 & 0 \\
0 & \cos\theta_{R} & \sin\theta_{R} \\
0 & -\sin\theta_{R} & \cos\theta_{R} \end{pmatrix}
\begin{pmatrix}
\sin\theta_{w} & 0 & \cos\theta_{w} \\
0 & 1 & 0 \\
\cos\theta_{w} & 0 & -\sin\theta_{w} \end{pmatrix}
\begin{pmatrix}
1 & 0 & 0 \\
0 & \cos\phi & -\sin\phi \\
0 & \sin\phi & \cos\phi \end{pmatrix}
\begin{pmatrix}
A \\
Z'\\
Z \end{pmatrix}\,,
\end{equation}
where $\tan\theta_{R}=g_{X}/g_{R}$, $\tan\theta_{w}=g'/g$, and $\phi\sim v^{2}/v_{R}^{2}$ ($v^{2}=2(k^{2}+k^{\prime\,2})$).

\subsection{$SU(2)_{1}\times SU(2)_{2}\times U(1)_{Y}$ model} \label{12 models}
We consider a model based on the gauge group $SU(2)_{1}\times SU(2)_{2}\times U(1)_{Y}$ \cite{Barger:1980p3439}, which is the simplest gauge extension of the SM containing a vector in the $(\mathbf{1},\mathbf{3})_{0}$ representation. The $SU(2)_{1}\times SU(2)_{2}$ symmetry is broken to the diagonal subgroup by the \textsc{vev} of a \mbox{bi-doublet} $\Delta \sim (2,2,0)$ 
$$
\left\langle\Delta\right\rangle \,= \begin{pmatrix}
f & 0 \\
0 & f
\end{pmatrix}\,.
$$
The $SU(2)_{L}$ gauge coupling is given by $1/g^{2}=1/g_{1}^{2}+1/g_{2}^{2}$. EWSB is accomplished by a doublet $H_{1}\sim (2,1,1/2)$ with \textsc{vev}
$$
\left\langle H_{1}\right\rangle\,=\frac{1}{\sqrt{2}}\begin{pmatrix}
0 \\
v \end{pmatrix}\,.
$$
The neutral mass matrix is diagonalized by the transformation
\begin{equation*}
\begin{pmatrix} 
W^{3}_{1} \\
W^{3}_{2} \\
B \end{pmatrix} = \begin{pmatrix}
\cos\theta_{L} & -\sin\theta_{L} & 0 \\
\sin\theta_{L} & \cos\theta_{L} & 0 \\
0 & 0 & 1 \end{pmatrix} \begin{pmatrix}
\sin\theta_{w} & 0 & \cos\theta_{w} \\
0 & 1 & 0 \\
\cos\theta_{w} & 0 & -\sin\theta_{w}
\end{pmatrix}
\begin{pmatrix}
1 & 0 & 0 \\
0 & \cos\theta' & -\sin\theta' \\
0 & \sin\theta' & \cos\theta' \end{pmatrix}
\begin{pmatrix}
A \\
Z' \\
Z \end{pmatrix}\,,
\end{equation*}
where $\tan\theta_{L}=g_{1}/g_{2}$, $\tan\theta_{w}=g'/g$ and $\theta'\sim v^{2}/f^{2}$ is the $Z$-$Z'$ mixing angle. On the other hand, mass mixing in the charged sector is diagonalized by the rotation
\begin{equation*}
\begin{pmatrix}
W_{1}^{+} \\
W_{2}^{+} \end{pmatrix}\,=
\begin{pmatrix}
\cos\alpha & -\sin\alpha \\
\sin\alpha & \cos\alpha \end{pmatrix}
\begin{pmatrix} 
W^{+} \\
W^{\prime\,+} 
\end{pmatrix}\,,
\end{equation*} 
where $\alpha\equiv \theta_{L}+\theta_{c}$, with $\theta_{c}\sim v^{2}/f^{2}$.

\bibliography{paper}{}
\bibliographystyle{JHEP}

\end{document}